\documentclass[nofootinbib,superscriptaddress, a4paper]{revtex4}

\usepackage{hyperref}
\usepackage{graphicx}
\usepackage[utf8]{inputenc}
\usepackage[T2A,T1]{fontenc}		
\usepackage{textcomp}
\usepackage{amsmath}
\usepackage{amsthm}
\usepackage{amssymb}
\usepackage{amsfonts}
\usepackage{mathrsfs}
\usepackage{bbm}
\usepackage{tabularx}
\usepackage{multirow}	
\usepackage{eurosym}
\usepackage{color}
\usepackage[dvipsnames]{xcolor}
\usepackage{fancyhdr}
\usepackage{fancybox}
\usepackage{stmaryrd} 
\usepackage{slashed}
\usepackage{dsfont}
\usepackage{lastpage}
\usepackage{refcount}
\usepackage{CJKutf8}
\usepackage{diagbox}

\newcommand{\me}{\mathrm{e}}
\newcommand{\mi}{\mathrm{i}}
\newcommand{\md}{\mathrm{d}}

\DeclareMathOperator{\trace}{tr}

\newcommand{\ket}[1]{{\left|#1\right\rangle}}
\newcommand{\bra}[1]{{\left\langle#1\right|}}

\renewcommand{\f}{\frac}

\renewcommand{\C}{{\mathbb C}}
\newcommand{\N}{{\mathbb N}}
\newcommand{\R}{{\mathbb R}}

\newcommand{\cE}{{\mathcal E}}

\newcommand{\cH}{{\mathcal H}}

\newcommand{\cN}{{\mathcal N}}

\newcommand{\cD}{{\mathcal D}}

\newcommand{\cS}{{\mathcal S}}

\renewcommand{\U}{\mathrm{U}}
\newcommand{\SU}{\mathrm{SU}}

\newcommand{\be}{\begin{equation}}
\newcommand{\ee}{\end{equation}}
\newcommand{\beq}{\begin{eqnarray}}
\newcommand{\eeq}{\end{eqnarray}}
\newcommand{\bes}{\begin{eqnarray}}
\newcommand{\ees}{\end{eqnarray}}

\renewcommand{\u}{{\mathfrak {u}}}
\newcommand{\su}{{\mathfrak {su}}}

\renewcommand{\u}{{\mathfrak{u}}}

\begin{document}

\title{Surface state decoherence in loop quantum gravity, a first toy model}

\author{{\bf Alexandre Feller}}\email{alexandre.feller@ens-lyon.fr}
		
\author{{\bf Etera R. Livine}}\email{etera.livine@ens-lyon.fr}
\affiliation{Univ Lyon, ENS de Lyon, Univ Claude Bernard, CNRS, Laboratoire de	Physique,	
  		F-69342 Lyon, France}
		
\begin{abstract}

The quantum-to-classical transition through decoherence is a major facet of the 
semi-classical analysis of quantum models that are supposed to admit a classical regime, as quantum gravity should be.
A particular problem of interest is the decoherence of black hole horizons and holographic screens induced by the 
bulk-boundary coupling with interior degrees of freedom. Here in this paper we present
a first toy-model, in the context of loop quantum gravity, for the dynamics of a surface geometry as an open quantum system at fixed total area.
We discuss the resulting decoherence and recoherence and compare the exact density matrix evolution to the commonly used  master equation approximation {\it \`a la} Lindblad underlining its merits and limitations.
The  prospect of this study is to have a clearer understanding of the boundary decoherence
of black hole horizons seen by outside observers.

\end{abstract}
		
\maketitle

\section{Introduction} 

Einstein's theory of gravitation, general relativity, describes the 
gravitational force as a manifestation of the curvature of space-time
while the matter and energy densities tell it how to curve. No 
background is assumed and every field is dynamical. One 
of the most fascinating prediction of general relativity is the formation of black 
holes in a gravitational collapse. The mathematical theory of black holes is very 
rich and led to causality studies, global techniques to analyze space-times,
to uniqueness and singularity theorems. What researchers came also
to understand is the holographic nature of gravity, the fact that volume degrees
of freedom can be described by some surface degrees of freedom. This led to the 
statement of the holographic principle \cite{Bousso_2002} which can serve as 
a guide for the understanding of the dynamic in quantum theories of gravity.
An attractive point of view on black holes is the membrane paradigm 
\cite{Damour_1982,Thorne_1986,Gourgoulhon_2005} which describes the horizon of
a black hole has a 2d surface living and evolving in 3d space. More recently, a 
similar point of view was developed for timelike surface \cite{Freidel_2013,Freidel_2015} 
and a dictionary with non equilibrium thermodynamics and the hydrodynamics of 
viscous bubbles was established. Thus, the fact that gravitation is about 
geometry and that Einstein's equation can be formulated in the 
language of a surface dynamic analogous to hydrodynamical equations
could serve as a new avenue to understand dynamical aspects of a
quantum theory of gravity such as loop quantum gravity.

Understanding gravity in the quantum regime is still an open issue and is 
the focus of active researches. Loop quantum gravity is a proposal of a 
quantum theory of gravity based on a non-pertubative canonical 
quantization of general relativity  (for textbooks, see \cite{Rovelli_book,
Rovelli_Vidotto_book,Thiemann_book}). The formalism is based on a $3+1$
formulation of general relativity in terms of the Ashtekar-Barbero 
connection and a densitized triad. The spectral analysis of geometrical 
operators such as the area of a surface or the volume of a region led
to the physical picture of a discrete space(-time) at the Planck scale. 
At the kinematical level, the natural basis of the state space is spanned 
by spin network states which are eigenstates of the area and volume 
operators. They live on a graph and are dressed with a spin on each
edge and intertwiner on each vertex carrying respectively area and 
volume information. In this context of quantum geometry, a surface $\cS$
is defined by the set of edges  of the spin network that it intersects, that is as the collection of the spins living on those edges, 
$\cS = \left( j_1, \dots, j_N \right)$. Its dynamic is controlled by the 
Hamiltonian constraint and all the other degrees of freedom of the universe.
This environment is composed for a closed surface of its exterior and interior, or more generally 
the rest of the spin network, and matter and field degrees of freedom. The direction
we want to particularly explore is the influence of an environment on the surface dynamic.

A careful analysis of the 
intertwiner space led recently to a new formulation of the phase space
of loop quantum gravity in term of spinors and the $U(N)$ 
formalism for an $N$-valent vertex \cite{Freidel_Livine_2010}. 
Interestingly, an $N$-valent vertex can be seen as a quantum polyhedron 
with $N$ faces and the $U(N)$ group appears to be the set of deformation 
preserving  the boundary area of this geometrical quantum object \cite{Livine_2014}.
The natural operators in this setting simply destroy a quantum of area in one 
place of the surface and recreate it at another. In a very straightforward
way, this point of view on quantum geometry allows to construct models
of the dynamic of a quantum surface and opens the way to the surface 
dynamic viewpoint on gravitation discussed above at the quantum level.

Alongside the comprehension of the quantum dynamic of gravity through the 
implementation of the Hamiltonian constraint at the quantum level
(for references and reviews on current researches, see  \cite{Thiemann_1996,
Bonzom:2011jv}), a major open question in loop quantum gravity is
its semi-classical regime and the recovery of general relativity.
A special focus is  devoted to understanding the renormalization
 of the theory by defining properly the coarse-graining of spin network states
(e.g. \cite{Livine_2014,Livine_Terno_2005,Dittrich:2013bza,Dittrich:2014ala,
Dittrich:2014wda}) and also by defining the proper notion of coherent
states of the quantum geometry \cite{Freidel_Livine_2011,Stottmeister:2015tua,Stottmeister:2015qxa,Stottmeister:2015vua}.
The emergence of 
a classical reality from a purely quantum one is an issue that dates back to 
the origin of quantum theory and is nowadays mostly understood thanks
to decoherence, a phenomenon recently observed  in (cavity) quantum 
electrodynamic and condensed matter experiments (for reviews and 
recent ideas, \cite{Haroche_book,Zurek_2003,Zurek_2008,Zurek_2009}). 
The heart of the idea behind decoherence is that every system is never
isolated but is an open quantum system in contact with an unmonitored
environment. Information about the quantum state of the system 
leaks inevitably in the environment through entanglement and this 
information remains lost to the observer. This leads in turn to the 
suppression of quantum superposition and interference effects into an 
effectively classical mixed state. The quantum states most robust 
to the ever monitoring of the environment are called pointer 
states and are the semi-classical states of the system.

What we propose to study in this paper is a first exploration of 
decoherence in loop quantum gravity by studying the dynamic of a 
quantum surface with $N$ patches. The total area, corresponding to the total  spin,
is supposed to be constant. As exposed above this system
is not isolated but in contact with an environment composed of all other
degrees of freedom available. Of course studying the surface dynamic in the full theory
would require to solve exactly the Hamiltonian constraint to obtain the 
true quantum dynamic which is out of reach. Still we know that the Hamiltonian couples the bulk and 
surface degrees of freedom. Instead, we construct effective models of the surface
interacting with an environment by using the $U(N)$ deformation operators
discussed above. The simplest, and quite general \cite{Feynman_Vernon_1963}, 
environment we can consider is a bath of harmonic oscillators coupled bilinearly to 
the area preserving deformation operators. Each deformation mode is then 
coupled to the environment. For this first inquiry, we limit
ourselves to the quantum measurement limit where the full dynamic is 
approximated to the interaction term only.
The first toy model we look at is a surface with two patches whose dynamic can 
be modeled has a spin, encoding the closure defect, whose three directions are coupled to harmonic oscillators.
This is a non trivial interaction seldom explored in studies on decoherence effect.
Interestingly, we obtain a decoherence phenomenon not on the value of the spin 
but only for certain quantum superposition of integer \emph{and} half-
integer spins. The decoherence time-scale appears to be independent of the spins 
on long time while the short time behavior maps the one studied using approximate 
methods. The decoherence factor decays exponentially with a decoherence time 
scaling as the relative distance between the spins. Those exact results contrasts
with master equation approaches which only capture short time behavior
and predict a decoherence as long as we have a quantum superposition 
of different spins. The physical origin of this difference comes from the 
model used for the environment which is, for Markovian equation, a memory-less
dynamical environment while it is considered non-dynamical for the exact toy 
model.

The paper is structured as follows. Section \ref{Surface geometry and quantization}
reviews the basic tools used in the analysis of the models. After some reminders
on decoherence in Section \ref{Decoherence and surface dynamic models}, 
Section \ref{Analysis of a toy model}  analyzes the exact behavior of the toy model of a 
two patches surface while Section \ref{Master equation approaches} deals with approximate 
master equation approaches of the complete dynamic. Section \ref{Conclusion}
concludes and open this discussion of surface state decoherence in loop quantum 
gravity.

\section{Surface geometry and quantization}
\label{Surface geometry and quantization}

\subsection{Spin as harmonic oscillators}

The geometry of a two dimensional surface $\mathcal{S}$ can be described from two
different point of view: the intrinsic one which relies on the Riemannian curvature
and the extrinsic one. The latter presupposes an embedding of the surface in a 
higher dimensional space like $\R^3$. The extrinsic curvature (also called second fundamental 
form) is defined as the variation of the surface normal vector $\mathbf{N}\in\R^{3}$ along the manifold $\mathcal{S}$. This normal vector also gives the integration measure on the manifold.This description is privileged by
 the canonical quantization of geometry in loop quantum gravity.
 
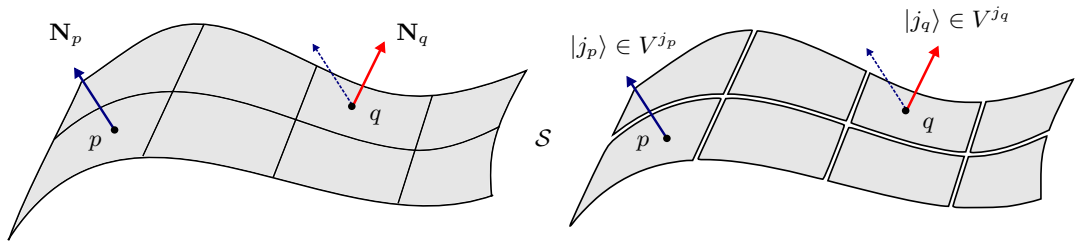
\begin{figure}[!ht]
 \label{extrins}
\begin{center}
	\begin{picture}(400,90)
		\put(0,0){\includegraphics[width=0.8\textwidth]{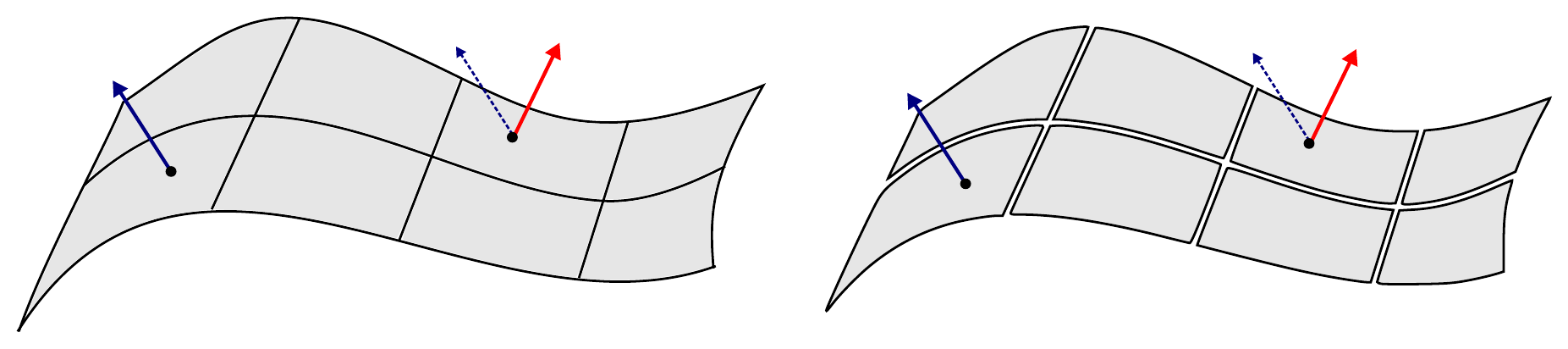}}		
		\put(202,40){$\cS$}
		\put(35,40){$p$} \put(240,40){$p$} \put(140,50){$q$} \put(347,47){$q$}
		\put(20,80){$\mathbf{N}_p$} \put(150,80){$\mathbf{N}_q$}
		\put(215,75){$\ket{j_p} \in V^{j_p}$} \put(340,85){$\ket{j_q} \in V^{j_q}$}
	\end{picture}
 \caption{Geometry of a 2d surface $\mathcal{S}$ in terms of the extrinsic curvature seen 
 		as the variation of the normal. The quantum theory describes $\mathcal{S}$ as 
 		a discretized set of patches $\cS_i$ whose quantum states live in $V^{j_i}$ 
 		a spin $j_i$ representation of $\SU(2)$.}
\end{center}
\end{figure}

The loop quantum gravity approach to the quantization of such a geometry is 
twofold. First we consider a discretization of $\mathcal{S}$ in terms of elementary 
surfaces (a face or a patch) $\mathcal{S}_i$. Each patch is defined by  its surface normal $\mathbf{N}_i\in\R^{3}$, whose norm is the area of the surface. It is further  provided with a phase 
space defined by a $\su(2)$ Poisson Bracket $\lbrace \mathbf{N}_i^{a},\mathbf{N}_i^{b} \rbrace
= \gamma\frac{8\pi G}{c^3} \epsilon^{abc}\mathbf{N}_i^{c} $.
This phase space is then canonically quantized to the operator commutator 
$[J_i^{a},J_i^{b}] = \gamma\frac{8\pi G \hbar}{c^3}\epsilon^{abc}J_i^{c}$. This is the 
basic postulate of loop quantum gravity. 
The proportionality 
factor has the dimension of an area and is related to the Planck area $G \hbar /c^3=l_{P}^{2}$, 
the only  dimensional quantity that appear in quantum gravity and $\gamma$ is the 
Immirzi parameter, a dimensionless number that fixes the scale of the theory.

The quantum state of 
each elementary surface patch $\mathcal{S}_i$ is then a vector of an irreducible $\SU(2)$ representation $V^{j_i}$. The spin $j_i$ then gives the area of that surface\footnotemark{} in Planck units $\gamma l_{P}^{2}$.
\footnotetext{
The area is classically given by the norm of the norm vector $|\mathbf{N}|=\sqrt{\mathbf{N}^2 }$. Since the normal vector $\mathbf{N}$ is quantized into the $\su(2)$ generator $\mathbf{J}$, the squared norm $\mathbf{N}^2$ becomes  the $\su(2)$ Casimir $\mathbf{J}^2$, whose spectrum is  $\sqrt{j(j+1)}$ in terms of the spin $j$.
So a traditional area spectrum in loop quantum gravity is $\sqrt{j(j+1)}$. However, taking a square-root naturally leads to non-polynomial observables and to quantization ambiguities. For instance, in the Schwinger representation of $\SU(2)$ representation in terms of harmonic oscillators, the norm $|\mathbf{N}|$ becomes a (quadratic) polynomial in the harmonic oscillator operators and has a unique consistent quantization as simply the spin $j$ \cite{Freidel_Livine_2010,Freidel_Livine_2011}. This is also the natural area spectrum when analyzing the $\SU(2)$-invariant observables and deformation algebra of intertwiners \cite{Livine_2013}.
}
The Hilbert space of a $N$ patches surface with fixed spins $j_1 \dots j_N$ is then 
\begin{align}
\cH_{j_1 \dots j_N} = V^{j_1} \otimes \cdots \otimes V^{j_N}
\end{align}
Intertwiners are defined as the $\SU(2)$-invariant subspace of this Hilbert space:
\be
{^{0}}\cH_{j_1 \dots j_N} = \textrm{Inv}_{\SU(2)}
\,\big{[}
V^{j_1} \otimes \cdots \otimes V^{j_N}
\big{]}
\ee
These singlet states are understood as the quantum counterpart of classical polyhedra 
\cite{Barbieri:1997ks,Bianchi:2010gc,Livine_2013}.
For the purpose of the article, we will focus on a surface with fixed area 
$A = \sum_{p=1}^{N} j_p$ with Hilbert space
\begin{align}
\cH_N^A = \bigoplus_{A = \sum_{p=1}^{N} j_p} \cH_{j_1 \dots j_N}\,.
\end{align}
Its $\SU(2)$-invariant subspace ${^{0}}\cH_N^A$ describes the Hilbert space  of the set of all polyhedron 
of area $A$. As it was shown in \cite{Freidel_Livine_2010}, these intertwiner spaces ${^{0}}\cH_N^A$ each 
carry an irreducible representation of the unitary group $\U(N)$, which can be be understood as the group of
deformation of quantum polyhedra at fixed total boundary area $A$. We will recall the definition of the $\u(N)$ 
generators below as the basic deformation operators for a quantum surface.

Finally, the total Hilbert space associated to a quantum surface $\cS$ with $N$ patches is 
\begin{align}
\cH_N = \bigoplus_{A\in \mathbb{N}} \cH_N^A
=\bigoplus_{\{j_p\}} \cH_{j_1 \dots j_N}
=\bigoplus_{\{j_p\}} V^{j_1} \otimes \cdots \otimes V^{j_N}
\,.
\end{align}

In this framework, studying the dynamic is naturally done through the study
of the deformations of the surface. In particular, the area of each face can 
evolve which means in the quantum theory changing the spin $j_e$ attached
to the face. The common $\SU(2)$ representation used in angular momentum theory
is not adapted for this purpose. But the Schwinger representation of the $\su(2)$ Lie algebra
in terms of harmonic oscillators is and we review its construction here
\cite{Girelli:2005ii,Freidel_Livine_2010}.

Let's focus on one spin (i.e. one elementary surface patch) and introduce two harmonic oscillators $a$ and $b$ whose 
commutation relations are naturally $[a,a^{\dagger}]=[b,b^{\dagger}]=1$. It is then 
straightforward to show that
\begin{align}
J_z = \frac{1}{2}(a^{\dagger}a - b^{\dagger}b)
\quad
J_{+} = J_{-}^{\dagger} =  a^{\dagger}b 
\end{align}
satisfy the $\su(2)$-algebra. The total energy of the oscillators $\mathcal{E} =
\frac{1}{2}(a^{\dagger}a + b^{\dagger}b)$ allows to write $\mathbf{J}^2 = 
\mathcal{E}(\mathcal{E}+1)$, so that the total energy gives exactly the spin 
$j$, i.e. the area of the elementary surface patch. Similarly, the energy difference corresponds to the magnetic quantum number 
$m$. The Hilbert space we are working with is then $\cH_{HO}\otimes\cH_{HO} = 
\oplus_{j} V^{j}$. Using standard notation, we have the correspondence 
between the spin and the harmonic oscillators states
\begin{align}
\ket{j,m} = \ket{n_a,n_b} \quad 
j=\frac{1}{2}(n_a + n_b) \quad 
m=\frac{1}{2}(n_a - n_b)
\end{align}
We can at once see that the action of $a$ or $b$ decreases the spin and thus the
area by $1/2$. The Schwinger representation admits natural operators allowing to move between different 
spin representation of the $\su(2)$-algebra, a feature more complicated to achieve with the 
standard representation.

Now consider a surface with $N$ faces described by spins $(j_i)_{i=1,...,N}$. We then 
indeed need $N$ pairs of harmonic oscillators $\left(a_i,b_i \right)_{i=1,...,N}$ to describe 
the surface state living in the Hilbert space $\cH_N = \cH_{HO}^{\otimes 2N}$. 
This representation naturally allows us to define a new set of 
operators that deform the surface. Following \cite{Girelli:2005ii,Freidel_Livine_2010}, we define the operator $E_{ij}$
that destroys a quantum of area at the face $j$ and creates one at $i$  by\footnotemark{}~:
\begin{align}
E_{ij} = a^{\dagger}_i a_j + b^{\dagger}_i b_j \quad
\end{align}
\footnotetext{We could 
define operator that only destroy or create quantum of area but we don't need them yet
for the present study. They are defined as $F_{ij} = a_ib_j-a_jb_i$ and their Hermitian conjugate $F_{ij}^{\dagger}$ \cite{Freidel_Livine_2011} and are used to define coherent intertwiner states.}
Clearly the action of those operators deforms the surface $\mathcal{S}$, preserve
the total area and are invariant under $\SU(2)$ rotations. So those operators act 
on each space $\cH_N^A $ without affecting the area $A$. The total area $A$ is related to the total energy of the 
oscillators by  $A = \sum_{p=1}^{N} j_p = \frac{1}{2} \sum_{p=1}^{N} E_{pp} $. The
operators $E_{ij}$ also satisfy the $\u(N)$ algebra \cite{Freidel_Livine_2010,Livine_2013}
\begin{align}
\left[E_{ij}, E_{kl}\right] = \delta_{jk}E_{il} - \delta_{il}E_{kj}
\end{align}
The group $\U(N)$ can thus be seen as the group of area preserving deformations
 of a discrete quantum surface  with $N$ faces. The (quadratic) Casimir operator $\mathcal{C}$ of this $\u(N)$ algebra is 
 \begin{align}
\label{Casimir_U(N)}
\mathcal{C}^{2} = \sum_{ij}E_{ij}^{\dagger}E_{ij} = 2A(A+N-2) + 2\mathbf{J}.\mathbf{J}
\end{align}
where $\mathbf{J} = \sum_{p=1}^N \mathbf{J_p}$ is the total spin operator. The operators $\mathbf{J}$ generate global $\SU(2)$ transformation on all spins simultaneously, corresponding to an overall 3d-rotation of the whole surface.
When this global $\SU(2)$ Casimir vanishes, $\mathbf{J}^{2}=0$, we are back on the $\SU(2)$-invariant subspace ${^{0}}\cH_N^A$ of quantum polyhedra. But in general, $\mathbf{J}^{2}$ is dubbed the ``closure defect'' \cite{Livine_2014,Charles_2016}.
This closure defect appears naturally when 
coarse-graining the spin network state. Nonetheless its physical significance is not yet 
perfectly understood but it is suspected to be related to curvature and torsion in the 
coarse-grained region induced be some quasi-local energy density.
Nevertheless, the important point for our concern is that the eigenstates of this operator $\mathbf{J}^{2}$ will by at the heart of our discussion of pointer states of the quantum surface  following decoherence.

Having now the kinematical scene for the system we want to understand and natural
operators to define its dynamic, we go on to discuss a very special class of states that play a central role
for the semi-classical understanding of the theory.

\subsection{Coherent states}

Coherent states play a very special role in the understanding of the quantum/classical 
transition in many areas of physics and also in quantum gravity. They allow to 
interpolate a classical geometry from its quantum description. $\SU(2)$
coherent states are the natural one to use in loop quantum gravity. Following \cite{Perelomov_1977},
a coherent states $\ket{j,g} $ is defined by applying an $\SU(2)$ rotation $g$ to a state analogous
to the vacuum in quantum optics that minimize the uncertainty relations such as the highest weight
state $\ket{j,m=j}$,
\begin{align}
\ket{j,g} = g\ket{j,j},\; g \in SU(2)
\end{align}
A key property of coherent states is that they remain coherent under the action of 
a $\SU(2)$-rotation. This follows directly from their very definition,
\begin{align}
h \ket{j,g} = \ket{j,hg}
\end{align}
Different ways exist to index coherent states. Instead of using the $\SU(2)$ rotation $g$, 
a coherent state can equivalently be labeled using spinors $z \in \C^2$. 
The highest weight vector is the spinor $\ket{\uparrow}$ and  can be mapped to any arbitrary unit spinor by  a rotation $g\in\SU(2)$, so that a $\SU(2)$ matrix contains the same information as a unit spinor $\langle z|z \rangle=1$. Explicitly the parametrization of $\SU(2)$ coherent states by spinors goes as:
\begin{align}
\ket{\uparrow}
=\ket{j=\f12,m=\f12}
=\begin{pmatrix} 1\\ 0 \end{pmatrix},
\quad
\ket{z} = \begin{pmatrix} z_0\\ z_1 \end{pmatrix}, \quad
g = \frac{1}{\sqrt{\langle z | z \rangle}} 
\begin{pmatrix}
z^0 & -\overline{z}^1 \\
z^1 & \overline{z}^0
\end{pmatrix}, \quad
g\ket{\uparrow} = \ket{z}\,,
\end{align}
\be
\ket{j,\uparrow}
=\ket{j,j}\,,
\quad
\ket{j,z}
=\Big{(}\sqrt{\langle z|z \rangle}\Big{)}^{2j}\,g\ket{j,\uparrow}\,.
\ee
Spinors for loop quantum gravity have been extensively  studied in \cite{Freidel_Livine_2010,Freidel_Livine_2011,Livine_2013}.
The explicit decomposition of a coherent states on the standard basis $\ket{j,m}$ used in 
angular momentum theory reads:
\begin{align}
\label{coherent_decompo}
\ket{j,z} = \sum_{m = -j}^{j} \sqrt{ 2j \choose j + m}
				(z^0)^{j+m} (z^1)^{j-m}
				\ket{j,m} \,.
\end{align}
Then the norm (and scalar product) between coherent states can  be calculated in terms of the simpler 
scalar product between spinors by the formula:
\begin{align}
\langle j',z' | j,z \rangle = \delta_{j'j}\langle z'|z \rangle^{2j}
\end{align}
Such coherent states are the basic tools for the construction of more interesting states such as
coherent intertwiner states or $U(N)$ coherent states for the semi-classical analysis of loop quantum gravity.

\section{Decoherence and surface dynamic models}
\label{Decoherence and surface dynamic models}

\subsection{About decoherence}

The destruction (or attenuation) of interference, called decoherence, of a quantum 
superposition through the entanglement of the system with an environment 
is at the heart of the modern understanding of the quantum to classical transition
(for reviews see \cite{Haroche_book,Zurek_2003,Zurek_2008}).
Decoherence comes from the leakage of information on the state of the system in an 
environment that can't be monitored by the observer. The states most immune to this 
constant monitoring of the environment, that entangle least with it, are called pointer
states and are in fact the natural classical states of the system. Pointer states are 
predictable and a quantum superposition of them evolves into a classical mixture. Pushed
even further, decoherence ideas are being used to understand more deeply the emergence
of a classical objective reality (Quantum Darwinism approach, \cite{Zurek_2009}).

Since the environment is unmonitored, the natural object to look at is the reduced 
density matrix of the system $\rho_\cS(t) = \trace_\cE{\left( U(t)\rho_{\cS\cE}(0)U^{-1}(t)
\right)}$ with dynamic ruled by $\mi \hbar \frac{\md \rho_\cS(t)}{ \md t} = 
\trace_\cE \left[ H,\rho_{\cS\cE}(t) \right]$. In most situations this equation 
cannot be solved exactly.

Their exist mostly two paths to analyze the open quantum dynamic ruled by the 
Hamiltonian $H = H_{\cS} + H_{\cE} + H_{\cS\cE}$, with $H_{\cS/\cE}$ the free
Hamiltonian and $H_{\cS\cE}$ the interaction term. The first method is the Feynman-Vernon path integral
approach \cite{Feynman_Vernon_1963}, an exact approach but difficult to manipulate
in general, and the second one is master equation approaches. Those equations have the advantage to 
be mathematically more accessible but rely on approximations which must be checked on the system
of interest for the results to be relevant. The most used approximations are the Born-Markov 
approximations which, simply stated, say that initially no correlations exist between the system
and the environment and that the environment has no memory (the correlation functions of the 
environment vanish on a timescale much smaller than any other dynamical or observational times).
A particular form of Markovian master equations is the Lindblad form. This subset of equations are the
most general form a quantum dynamic can take (constrained by positivity and complete positivity of 
the reduced dynamic). Those different approaches will by used and compared in this paper 
for the surface dynamic we are interested in.

\subsection{The general model}

We are interested in the dynamic of a quantum surface $\cS = \left(j_1 \dots j_N \right)$ whose 
total area $A= \sum_{p=1}^{N} j_p$ is supposed to be a constant of motion having in mind a black
hole at equilibrium. We recall that in loop quantum gravity $\cS$ is just a part of a spin network state and 
the remaining degrees of freedom will here be considered as its environment. The Hilbert space $\mathcal{H_{\cS}}$
we work with is
\begin{align}
\mathcal{H_{\cS}} = \bigoplus_{A = \sum_{p=1}^N j_p}
\left( \bigotimes_{p=1}^N V^{j_p} \right)
\end{align}
Using the deformation formalism, we can use the operators $E_{ij}$ to
construct a natural interaction $H_{\cS\cE}$ between $\cS$ and $\cE$ where the environment excites 
each deformation $E_{ij}$ through an operator $V_{ij}$ so that 
\begin{align}
\label{interaction}
H_{\cS\cE}= \sum_{i,j = 1}^N E_{ij} \otimes V_{ij}
\end{align}
Such an interaction can be seen to emerge from the structure of the 
spin network by remembering that the graph encodes relationships.    
Hermicity requires that $V_{ij}^{\dagger} = V_{ji}$. For now, we don't specify the explicit form of those
operators. Nonetheless the typical environments considered in decoherence studies resume to a bath 
of harmonic oscillators which we will suppose in the remaining of the paper. For instance, those harmonic 
oscillators could model any matter fields from the bulk or the exterior of a black hole. From the Hawking
radiation, thermal states for the environment would be the most natural. The free Hamiltonian 
$H_{\cE}$ of the environment is thus the energy of a set of harmonic oscillators. Concerning the free
dynamic of the system $H_{\cS}$ we suppose that it has a contribution proportional to the area of 
the surface so that $H_{\cS} = \sum_{i=1}^N E_{ii}$. Since the area is fixed in our problem, 
such a term has no contribution to the global dynamic. Of course, it would be natural
to include higher order contribution of the $E_{ii}$ operators (as for instance a Bose-Hubbard type term
\footnote{To remind the reader, the Bose-Hubbard Hamiltonian used in cold atom physics is of the from 
$H = -t\sum_{\langle i,j \rangle} b^{\dagger}_i b_j + b^{\dagger}_j b_i + \frac{U}{2}\sum_i n_i(n_i+1)
-\mu \sum_i n_i $ where $t$ is a hopping constant, $U$ the interaction term and $\mu$ the chemical potential.
A Bose-Hubbard model for a quantum black hole would be in our setting $H= -t\sum_{\langle i,j \rangle} E_{ij}
+\frac{U}{2}\sum E_i(E_i +1)$. Since the $U(N)$ generators are now composed of two different
species $a$ and $b$ instead of a single one $b$, the physics of this model remains to 
be understood.}) but we leave this analysis for future works.

\begin{figure}[!ht]
\label{surface}
\begin{center}
	\begin{picture}(200,130)
		\put(0,0){\includegraphics[width=0.4\textwidth]{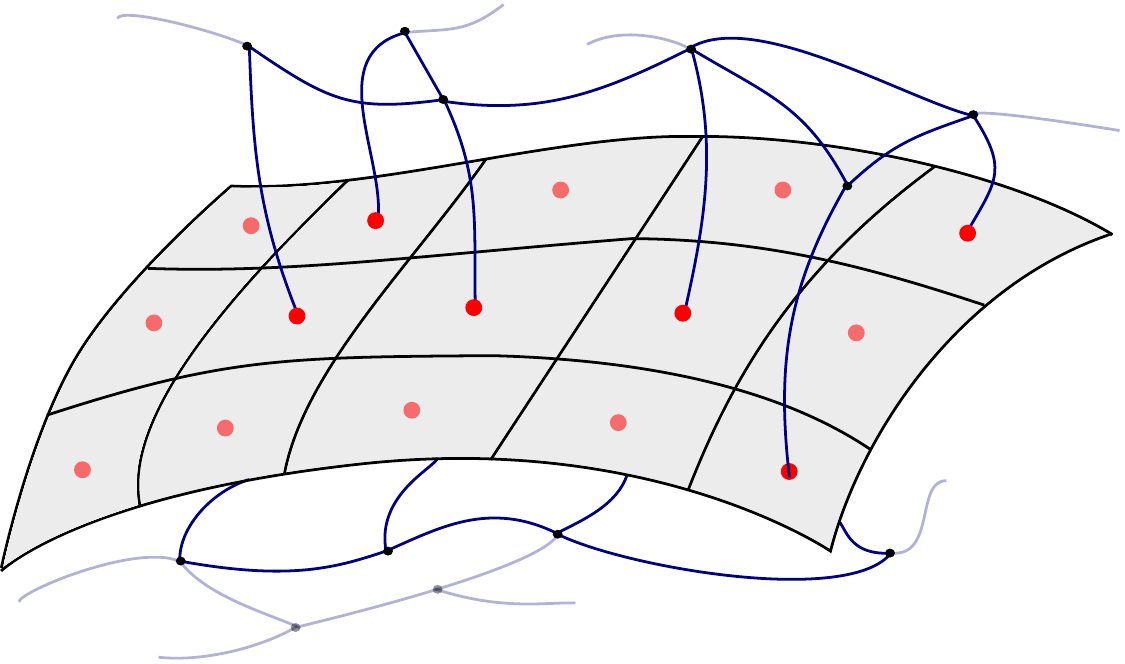}}		
		 \put(170,40){$\cS$}
	\end{picture}
 \caption{A surface $\cS$ is defined as a subset of a spin network while its environment is 
 		the remaining. The origin of the interaction between the patches comes from the 
 		structure of the graph.}
\end{center}
\end{figure}

\subsection{Decoherence from master equation}

Having now set the dynamic our quantum surface, we would like to understand the influence 
the environment has on the evolution of states of the system, especially if decoherence occurs
for certain privileged states or geometrical quantities that could then be labeled has 
pointer states of the surface.

Following the traditional route of master equation to tackle those problems, we propose
to study the surface open quantum dynamic through a Lindblad master equation with 
the deformation operators $E_{ij}$ as jump operators. This is a natural choice in the  light
of the dynamic \eqref{interaction} where the deformation modes are excited by the environment.
\begin{align}
\frac{\md\rho_{\mathcal{S}}}{\md t}
= \sum_{i,j =1}^{N}
	E_{ij}\rho_{\mathcal{S}}E_{ij}^{\dagger}
	- \frac{1}{2}
	\left( 
	E_{ij}^{\dagger}E_{ij} \rho_{\mathcal{S}} + \rho_{\mathcal{S}}E_{ij}^{\dagger}E_{ij}  
	\right)
= \sum_{i,j =1}^{N} 
	E_{ij}\rho_{\mathcal{S}}E_{ij}^{\dagger}-\frac{1}{2}\left( \mathcal{C}^{2}\rho_{\mathcal{S}}  
	+ \rho_{\mathcal{S}} \mathcal{C}^{2} \right)
\end{align}

To grasp the potential decoherence induced by such a coupling, we can look at the evolution induced 
by this equation on an initial quantum superposition of highest weight states $\ket{\psi_J^A}$ 
verifying $E_1\ket{\psi_J^A} = (A+J)\ket{\psi_J^A}$, $E_2 \ket{\psi_J^A} = (A-J)\ket{\psi_J^A}$
and $E_{ki}\ket{\psi_J^A} = 0$ for $k<i$ with $A$ and $J$ respectively representing 
the area of the surface and its closure defect. They are the $U(N)$ analogue of the $\ket{j,j}$ state of 
$\SU(2)$. The short time evolution of the purity of the coherence of the reduced density
matrix $\trace{\left( \rho_{JJ'}\rho_{J'J} \right)}$ with initial state $\frac{\ket{\psi_J^A} + \ket{\psi_{J'}^A}}{\sqrt{2}}$
can be directly obtained 
\begin{align}
\left. \frac{\md}{\md t}\trace{\left( \rho_{JJ'}\rho_{J'J} \right)} \right|_{t=0} = 
- \left[ A(N-2) + J + J' + (J-J')^2 \right]  
\trace{\rho_{JJ'}\rho_{J'J}}
\end{align}
The damping factor is always positive and composed of three terms. The last one $(J-J')^2$ is the 
one we where looking for which induced a decoherence effect on quantum superposition of
geometry with different closure defect and leads to a superselection rule on it. The second
shows that the state most immune to entanglement with the environment is the geometry without
defect while the first shows that the greater the area is the more entangled the system will be.

Nonetheless, many questions remain to be clarified from this naive discussion. What we will see
in this paper from exact and approximate special cases deduced from the previous model is that 

\begin{itemize}
\item The Markovian hypothesis must be discussed and its validity clarified. We will see that for non 
	dynamical environment, this hypothesis is jeopardized for a complete description at all 
	times but is only correct on short timescales.
\item Depending on the environment, a recoherence is not excluded for superposition of states 
	with different values of the defect and cannot be hinted with a short time approximation analysis
	(the roots of this behavior are to be sought in the compactness of $\SU(2)$ group).
	 This phenomena disappears when a large and dynamical environment is considered. 

\end{itemize}

\section{Analysis of a toy model}
\label{Analysis of a toy model}

In this section, we study the open dynamic of a toy model of a quantum surface with 
two faces and focus on decoherence effect. The end goal is to have a clear 
understanding of the long time behavior of the system, exhibit the pointer
states and their physical significance. Those steps will serve as the basis 
for the analysis of a more realistic model of the open surface dynamic in 
quantum gravity. We limit our exact study to the measurement limit by 
neglecting the free dynamic of the environment. Thus the domain of validity 
of the following results are in fact limited to timescales smaller then any 
dynamical times of the environment.

\subsection{Motivation}

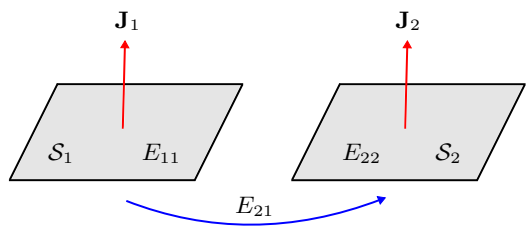
\begin{figure}[!ht]
\begin{center}
	\begin{picture}(200,100)
		\put(0,0){\includegraphics[width=0.4\textwidth]{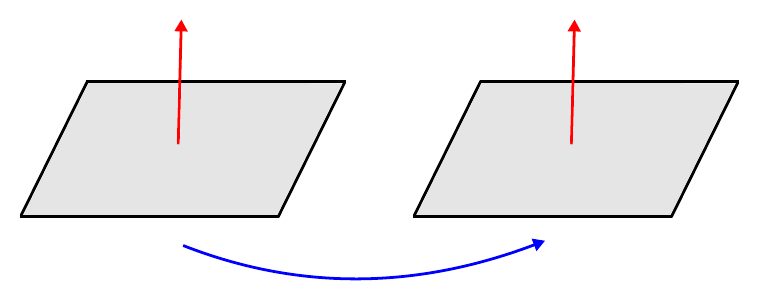}}		
		\put(20,30){$\cS_1$} \put(165,30){$\cS_2$}
		\put(90,10){$E_{21}$} \put(55,30){$E_{11}$} \put(130,30){$E_{22}$}
		\put(45,80){$\mathbf{J}_1$} \put(150,80){$\mathbf{J}_2$} 
	\end{picture}
 \caption{The dynamic of the two patches $(\cS_1,\cS_2)$ toy model is encoded 
 		in the deformation operator $E_{21}$. In loop quantum gravity, the kinematics 
 		of a surface is encoded in its  		normal $\mathbf{J}$ whose norm is 
 		the area of the surface and are found here in the operators $E_{11}$ and 
 		$E_{22}$. 
\label{N2_fig}}
\end{center}
\end{figure}

From the interaction $H_{\cS\cE}= \sum_{i,j = 1}^N E_{ij} \otimes V_{ij}$ we can motivate the introduction
of the toy model by looking explicitly at the $N=2$ patches model, see fig. \ref{N2_fig}. We work in the subspace 
of $\cH_2 = \cH_{HO}^{\otimes 4}$ with fixed area. By introducing the operators 
\begin{align}
L_z= \frac{E_{11}-E_{22}}{2}, \quad L_+ = E_{12}, \quad L_- = L_+^{\dagger}
\end{align}
we can rewrite the interaction in the form  
\begin{align}
H_{\cS\cE} &= \left( L_z \otimes \frac{V_{11}-V_{22}}{2} \right) + 
		   \left( L_x \otimes \frac{V_{12}+V_{21}}{2} \right) + 
		   \left( L_y \otimes \frac{V_{21}-V_{12}}{2\mi} \right) +
		   \left( \frac{E_{11}+E_{22}}{2} \otimes \frac{V_{11}+V_{22}}{2} \right)
\end{align}
Using those definitions, we can check the form of the $U(N)$ Casimir operator \eqref{Casimir_U(N)}
explicitly and obtain with the notation of this section that $\sum_{ij}E_{ij}^{\dagger}E_{ij} = \frac
{E^2}{2} + 2\mathbf{L}.\mathbf{L}$ with $E$ the total energy of the oscillators describing the patches.
Thus we have $\mathbf{L}^2 = \mathbf{J}^2$ and the eigenvalues of $\mathbf{L}$ correspond exactly
to the closure defect of the surface. Nonetheless, $\mathbf{L}\ne\mathbf{J}$ except in the special 
case where the spins are decoupled.

For concreteness, we choose in the remaining the environment to be a bath of harmonic oscillators.
We will look at the case of only one oscillator at first and then generalize to an arbitrary number.
So the model Hamiltonian we consider is 
\begin{align}
\label{N2_hamiltonian}
H_{\cS\cE} = \mathbf{L} \otimes \mathbf{p} +
			\left( \frac{E_{11}+E_{22}}{2} 
			\otimes \frac{V_{11}+V_{22}}{2} \right)
\end{align}
We then have a dynamic with the three directions of a spin coupled to the environment with an 
additional coupling to the energy of the oscillators and the spin. Since the area (i.e. the total energy 
of the oscillators) is fixed, the second term is non-dynamical\footnotemark{}. The first term of this interaction is the non-trivial one
and involves three non-commuting observables coupled to the environment.
\footnotetext{
If we had not fix the area, this second term in $(E_{11}+E_{22})\otimes (V_{11}+V_{22})$ would very likely imply a decoherence of quantum superposition of the area, and thus lead to a classical notion of surface area at late time.
}

The program of this section is first to look at the potential decoherence effect induced by 
the interaction $\mathbf{L} \otimes \mathbf{p}$ with a single oscillator as the environment and then
explore the consequences of a large environment.

\subsection{Reduced density matrix}

We focus now on the study of the spin part of the interaction $H_{\cS\cE} = \mathbf{L} \otimes \mathbf{p}$
with a single mode environment and want to understand the decoherence it induces. We look at a possible 
decoherence on the value of the total spin  $j$ which is the quantum number associated to the operator
 $\mathbf{L}^2$, the same $\mathbf{L}$ that  appears in the interaction. Since we have in mind the dynamic of the 
 horizon of a black hole, we naturally consider our system to be the spin. On the other hand, it would be also legitimate
 to reverse the problem and focus on the induced dynamic of the oscillator describing the exterior observer. This is what is 
 develop in appendix \ref{Density_matrix_env} for the toy model of this section.
 
Going back to the surface, the idea is then to study the evolution of a superposition of coherent states of the system
\begin{align}
\ket{\psi} = \frac{1}{\sqrt{2}} \left( \ket{j,g} + \ket{j',g'} \right)
\end{align}
The initial state of the harmonic environment is supposed to be the vacuum
\footnotetext{Considering a thermal state for the environment would be more 
accurate from our knowledge of Hawing radiation. But since in this first
investigation $H_\cE = 0$, this case is not relevant.
}
and uncorrelated to the state of the system so that the global initial state is
\begin{align}
\label{superposition}
\ket{\psi_{\cS\cE}} = \ket{\psi} \otimes \ket{0}
\end{align}

The evolution of this state is obtained most simply by developing the vacuum state on the impulsion
basis of the environment $\ket{0} = \int_{\R^3} \phi(\mathbf{p}) \ket{\mathbf{p}} \; \md \mathbf{p}$
where $\phi(\mathbf{p}) \propto \me^{-\lambda p^2} $ is a Gaussian wave function 
(the parameter $\lambda$ is not specified explicitly for it will be easier to obtain more general results keeping it)
since we have
$U(t)\ket{j,g}\ket{\mathbf{p}} = \ket{j, \me^{-\frac{\mi t }{\hbar}\frac{\boldsymbol{\sigma}.\mathbf{p}}{2}} g}
\ket{\mathbf{p}}$. The system remains in a coherent state when the environment is in an eigenstate of the
momentum operator.

The central object we want to calculate is the reduced density matrix of the system $\rho_{\cS}(t) = \trace_{\cE}
\left( {U(t) \rho_{\cS\cE}(0) U^{-1}(t)} \right)$ which characterizes completely the dynamic of the system alone.
Decoherence effects will be seen by analyzing the long time evolution of the $j \ne j'$ matrix elements and by showing
that they tend to zero. By introducing projection operators $P_j$ on the subspace of spin $j$, we can then focus
on certain elements of the reduced density matrix $\rho_{\cS}^{jj'}(t) = P_j \rho_{\cS} (t) P_{j'}$. Those operators
$\rho_{\cS}^{jj'}(t)$ contain all the information about the coherence between superposition of spins $j$ and $j'$.
\begin{align}
\rho_{\cS}^{jj'}(t) &= \frac{1}{\mathcal{N}'}
				\trace_{\cE}\left[ 
				\int_{\R^3 \times \R^3}
					\ket{j, \me^{-\frac{\mi t }{\hbar}\frac{\boldsymbol{\sigma}.\mathbf{p}}{2}} g}
					\bra{j', \me^{-\frac{\mi t }{\hbar}\frac{\boldsymbol{\sigma}.\mathbf{q}}{2}} g'}
					\otimes
					\ket{\mathbf{p}}\bra{\mathbf{q}}
					\me^{-\lambda\left(p^2 + q^2 \right)}
					\; \md \mathbf{p} \md \mathbf{q} 
					\right] \nonumber \\
			&= \frac{1}{\mathcal{N}'}
				\int_{\R^3} 
				\ket{j, D\left( \frac{tp}{2\hbar}, \hat{p}\right) g}
				\bra{j', D\left( \frac{tp}{2\hbar}, \hat{p}\right) g'}
				\me^{-2\lambda p^2}
				\; \md \mathbf{p}
\end{align}
where $D(\theta,\hat{n})= \me^{-\mi\theta \boldsymbol{\sigma}.\hat{\mathbf{n}}} 
= \cos(\theta) - \mi \boldsymbol{\sigma}.\hat{\mathbf{n}} \sin(\theta)$ and 
$\cN' = 2\left(2\lambda/\pi)\right)^{1/2}$ is the normalization constant. The mathematical details 
are a bit cumbersome and before delving into them we focus on particular case that highlights 
the general results.

\subsection{A simple calculation: decoherence for $j'=0$}

We are going to calculate the reduced density matrix in the simple case where $j'=0$ and the coherent state
is just $\ket{j,j}$. This last restriction can be lifted to any coherent state $\ket{j,g}$ by simply choosing the 
proper spherical coordinates in the following calculations.  
\begin{align}
\rho_{\cS}^{j}(t) = \int_{\R^3} 
				D\left( \frac{tp}{2\hbar}, \hat{p}\right) \ket{j,j}
				\me^{-2\lambda p^2}
				\; \md \mathbf{p}
\end{align}
The overall normalization factor $\cN'$ is not written explicitly and $\hbar = 1$ here. Developing explicitly the action 
of the rotation operator, we have according to formula \eqref{coherent_decompo}
\begin{align}
\rho_{\cS}^{j}(t) &= \int_{\R^3} 
				\sum_{m = -j}^{j} \sqrt{ 2j \choose j + m}
				(z^0_p)^{j+m} (z^1_p)^{j-m}
				\ket{j,m}
				\me^{-2\lambda p^2}
				\; \md \mathbf{p} \nonumber \\
			&\text{with }\; z^0_p = \cos\left( \frac{tp}{2} \right) - \mi \sin\left( \frac{tp}{2} \right)\cos(\theta)
		 	   \;\text{ and }\;  z^1_p  = -\mi \sin(\theta)\me^{\mi \phi}
\end{align}
with $\left( \theta, \phi \right)$ the spherical coordinates. We now proceed to the explicit calculation
of the integrals in this coordinate system with the measure $\md \mathbf{p} = p^2 \sin(\theta) \md \theta \md \phi \md p$.
The integral over the $\phi$ angle leads to a $2 \pi \delta_{j-m,0}$ contribution. The integral over the angle $\theta$ is then straightforward to do\footnotemark, leaving us with an integral over the norm $p$:
\begin{align}
\rho_{\cS}^{j}(t) = 2 \pi \int_0^{\infty}	
			\frac{2}{2j+1} \frac{\sin\left( (2j+1) \frac{tp}{2}\right)}{\sin\left( \frac{tp}{2} \right)}
			\me^{-2\lambda p^2} 
			 p^2 \: \md p \;
			 \ket{j,j} 
			= \sum_{m=-j}^{j} \frac{4 \pi}{2j+1} 
		\int_0^{\infty}	
			\me^{2 \mi m tp}
			\me^{-2\lambda p^2} 
			 p^2  \md p \;
		\ket{j,j}
\end{align}
\footnotetext{
The first integral over  the $\phi$ angle gives:
$$
\rho_{\cS}^{j}(t) = 2 \pi \int	
			\left(  \cos\left( \frac{tp}{2} \right) - \mi \sin\left( \frac{tp}{2} \right)\cos(\theta) \right)^{2j}
			\ket{j,j} \me^{-2\lambda p^2} 
			 p^2 \sin(\theta) \md \theta \md p\,.
$$
The remaining integral over the angle $\theta$ is a trigonometric integral:
$$
\int_0^{\pi} \left(  \cos\left( \frac{tp}{2} \right) - \mi \sin\left( \frac{tp}{2} \right)\cos(\theta) \right)^{2j} 
		\sin(\theta) \;\md \theta 
		= \frac{2}{2j+1} \frac{\sin\left( (2j+1) \frac{tp}{2}\right)}{\sin\left( \frac{tp}{2} \right)}
		= \frac{2}{2j+1} \sum_{m=-j}^{j} \me^{2 \mi m tp}\,.
$$
}
%
%
This last step properly highlights  the decoherence effect. Indeed, we have a sum over modes of the Fourier transform of a Gaussian distribution, implying a Gaussian decay for all modes except for the zero-mode. This zero-mode gives the remaining coherence of our quantum state at late time $t\rightarrow+\infty$.

Now more explicitly, we have that 
\begin{align}
\sum_{m=-j}^{j} \me^{2 \mi m tp}
= \left\{ 
	\begin{array}{l l}
  		2 \sum_{m=1/2}^{j} \cos(2mtp) & \quad \text{if } j \in \N+1/2 \\
 		1 + 2\sum_{m=1}^{j} \cos(2mtp)  & \quad \text{if } j \in \N 
  	\end{array} 
	\right.
\end{align}
clearly leading to a non-zero limit for integer spins:
\begin{align}
\rho_{\cS}^{j}(t) 
	&= \left\{ 
		\begin{array}{l l}
  			\frac{8\pi}{2j+1}\sum_{m=1/2}^{j} \int_0^{\infty} p^2 \cos(2mtp) \me^{-2\lambda p^2} \: \md p 
  			\; \ket{j,j}    
  			& \quad \text{if } j \in \N+1/2 \\
 			\frac{4\pi}{2j+1}
 				\left( 
 				\rho^0_{\cS } +2 
 				\sum_{m=1}^{j} \int_0^{\infty} p^2 \cos(2mtp) \me^{-2\lambda p^2} \: \md p  
 				\right)
 			\; \ket{j,j}    
 			& \quad \text{if } j \in \N
  		\end{array} 
		\right.
\end{align}
where all the integrals can be evaluated exactly:
\be
\int_0^{\infty} p^2 \me^{-\lambda p^2} \: \md p  = \frac{\sqrt{\pi}}{4\lambda^{3/2}}\,
\quad
\int_0^{\infty} p^2 \cos(mtp) \me^{-\lambda p^2} \: \md p  = 
\frac{\sqrt{\pi}}{8} \frac{1}{\lambda^{3/2}} \left( 2 - \frac{k^2 t^2}{\lambda} \right) \me^{-\frac{k^2t^2}{4\lambda}}\,.
\ee
Beside the zero-mode, all the remaining integrals all tend to zero at infinity.
Let's us not forget that we have omitted from the beginning the global normalization of the states to simplify the 
equations. The conclusion from this simplified version of the reduced density matrix shows that we cannot expect a full 
decoherence on the spin. Only coherence between half-integer spins is suppressed as time goes to infinity. 
Coherence with integer spins still exists and the limit value of the reduced density matrix is 
(up to a normalization factor $4\pi\frac{\sqrt{\pi}}{4\cN' \lambda^{3/2}}$)
\begin{align}
\label{limit_j'_0}
\rho_{\cS}^{j}(t) \underset{t \to \infty}{\longrightarrow} \frac{1}{2j+1} \,.
\end{align}
Nevertheless, apart from this zero-mode contribution, all the other modes lead to Gaussian decay in $\me^{-\frac{k^2t^2}{4\lambda}}$ in terms of the original Gaussian width $\lambda$ and the Fourier mode $k$. So we see a clear decoherence except for that remaining limit coherence decreasing with the spin $j$.

\subsection{Bath of harmonic oscillators} 

The previous calculations we did were with a single harmonic oscillator as an environment. We now consider a 
bath of $N$ harmonic oscillators and analyze the consequences on the dynamic of our system. In fact we show that 
only the time scales has changed by the presence of a bath but the mathematical expressions of the reduced density 
matrix are mostly the same as those already obtained.

The interaction is then $H_{\cS\cE} = \gamma \mathbf{L} \otimes \sum_k \mathbf{p}_k$ where $\gamma$ is a coupling constant. 
We again look at the evolution of the same initial state with the vacuum $\ket{0}$ being the vacuum of the whole bath
\begin{align}
\ket{\psi_{\cS\cE}} =\frac{1}{\sqrt{\cN}} \left( \ket{j,g} + \ket{j',g'} \right) \otimes \ket{0}
\end{align}
We again drop the normalization in the following. Using the same method we have 

\begin{align}
\rho_{\cS}^{jj'}(t ; N) &= \trace_{\cE}\left[ 
				\int_{\R^3 \times \R^3}
					\ket{j, \me^{-\frac{\mi t \gamma }{\hbar}\frac{\boldsymbol{\sigma}. \sum_k \mathbf{p}_k}{2}} g}
					\bra{j', \me^{-\frac{\mi t \gamma }{\hbar}\frac{\boldsymbol{\sigma}. \sum_k \mathbf{q}_k}{2}} g'}
					\otimes
					\ket{\mathbf{p}}\bra{\mathbf{q}}
					\me^{-\lambda\left(\sum_k p_k^2 + q_k^2 \right)}
					\; \md \mathbf{p} \md \mathbf{q} 
					\right] \nonumber \\
			&= \int_{\R^3} 
				\ket{j, \me^{-\frac{\mi t \gamma}{\hbar} \boldsymbol{\sigma}. \sum_k \mathbf{p}_k} g}
				\bra{j',\me^{-\frac{\mi t \gamma}{\hbar} \boldsymbol{\sigma}. \sum_k \mathbf{p}_k} g'}
				\me^{-2\lambda \sum_k p_k^2}
				\; \md \mathbf{p}
\end{align}
All that matters here is the center of mass dynamic $\sum_k \mathbf{p}_k$ in which the 
Gaussian width will be affected. To achieve this we perform a change of variable
 (with a Jacobian equal to one)
\begin{align}
\begin{pmatrix}
\mathbf{p}_1 \\ \vdots \\ \vdots \\ \mathbf{p}_N
\end{pmatrix} = 
\begin{pmatrix}
1 & -1 \cdots & -1 \\
0 & \ddots & 0 \\
0 & 0 & 1
\end{pmatrix}
\begin{pmatrix}
\mathbf{p}_G \\ \mathbf{u}_1 \\ \vdots \\ \mathbf{u}_{N-1}
\end{pmatrix}
\end{align}
We can then perform the Gaussian integral over the $\mathbf{u}_i$ variable without any issues. This gives again 
a Gaussian contribution of the form $\frac{2\pi^{(N-1)/2}}{(N2^{N-1})^{1/2}}\me^{\frac{N-1}{N}\mathbf{p}_G^2}$.
Finally 
\begin{align}
\rho_{\cS}^{jj'}(t ; N) = \sqrt{\frac{\pi^{N-1}}{N}} \int_{\R^3} 
				\ket{j, \me^{-\frac{\mi t \gamma}{\hbar} \boldsymbol{\sigma}. \mathbf{p}} g}
				\bra{j',\me^{-\frac{\mi t \gamma}{\hbar} \boldsymbol{\sigma}. \mathbf{p}} g'}
				\me^{-2\frac{\lambda}{N} p^2}
				\; \md \mathbf{p}
\end{align}
We can then deduce the scaling law satisfied by the reduced density matrix as a function of $N$. By 
comparing the above formula by the one with one oscillator, reinserting the normalization, we have 
\begin{align}
\label{densitymatrix_bath}
\rho_{\cS}^{jj'}(t ; N) = \frac{1}{N^2}\left(\frac{\pi^2}{2\lambda}\right)^{\frac{N-1}{2}} 
				\rho_{\cS}^{jj'}(\gamma \sqrt{N}t)
\end{align}
We conclude that in presence of $N$ harmonic oscillators instead of only one, the decoherence 
timescale is shorten by a factor $\sqrt{N}$. Only the convergence speed is affected, not 
the shape of the decoherence factor.

\subsection{Studying the general spin case}

The general case is more involved mathematically. We again focus on the initial coherent state with $g=\begin{pmatrix}
1 \\ 0 \end{pmatrix}$ and suppose that $j'>j$. To understand the long time behavior of the projected reduced density matrix, we look at
its norm first
\begin{align}
\label{fullequation}
\trace{\rho_{\cS}^{jj'}(t) \rho_{\cS}^{j'j}(t)} 
	&= \int_{\R^3 \times \R^3} 
	 	\langle  D\left( tq/2\hbar, \hat{q}\right) g |  D\left( tp/2\hbar\right) g \rangle ^{2j}
		\langle  D\left( tp/2\hbar\right) g' |  D\left( tq/2\hbar\right) g' \rangle ^{2j'}				
		\me^{-2\lambda  (p^2 + q^2)}
			\; \md \mathbf{p} \; \md \mathbf{q} \nonumber \\
	&= \int_{\R^3 \times \R^3} 
		\left( \overline{z}_q^0 z_p^0 + \overline{z}_q^1 z_p^1 \right)^{2j}
		\left( \overline{z}_p^0 z_q^0 + \overline{z}_p^1 z_q^1 \right)^{2j'}
		\me^{-2\lambda  (p^2 + q^2)}
			\; \md \mathbf{p} \; \md \mathbf{q} \nonumber \\
	&=\sum_a^{2j}\sum_b^{2j'} {2j \choose a}{2j' \choose b}
		\int_{\R^3 \times \R^3} 
		\left( \overline{z}_q^0 z_p^0 \right)^{2j-a}\left(\overline{z}_q^1 z_p^1 \right)^{a}
		\left( \overline{z}_p^0 z_q^0 \right)^{2j'-b} \left(\overline{z}_p^1 z_q^1 \right)^{b}
		\me^{-2\lambda  (p^2 + q^2)}
		\; \md \mathbf{p} \; \md \mathbf{q} 
\end{align}
We perform the explicit calculation in spherical coordinates again. The integral over the azimuthal angles
$\phi_p$ and $\phi_q$ give  a delta function contribution 
$\int_0^{2\pi} \me ^{\mi(\phi a - \phi b)} \; \md \phi  = 2 \pi \delta(a-b)$, so 
\begin{align}
\trace{\rho_{\cS}^{jj'}(t) \rho_{\cS}^{j'j}(t)} 
	= (2\pi)^2 \sum_a^{2j}{2j \choose a}{2j' \choose a}
		&\int_{\R^3 \times \R^3} 
		\left( \overline{z}_q^0 z_p^0 \right)^{2j-a} \left( \overline{z}_p^0 z_q^0 \right)^{2j'-a}
		\left( \sin(\theta_p)\sin(\theta_q)\sin\left(\frac{pt}{2}\right)\sin\left(\frac{qt}{2}\right) \right)^{2a} \nonumber\\
		& p^2 q^2 \sin(\theta_q)\sin(\theta_p)
		\me^{-2\lambda  (p^2 + q^2)}
		\; \md \mathbf{p} \; \md \mathbf{q} 
\end{align}
We now look at the integral over the polar angles $\theta$. First we have 
\begin{align}
\left(z_p^0 \right)^{2j-a} = \sum_{k=0}^{2j-a} {2j-a \choose k} 
					\cos^{2j-a-k} \left(\frac{pt}{2}\right)
					\sin^k\left( \frac{pt}{2} \right)
					\mi^k \cos^k\left(\theta_p\right)
\end{align}
Gathering all the terms containing the angle $\theta_p$, we need to evaluate an integral 
of the form
\begin{align}
\int_0^{\pi} \cos^{k+q}(\theta) \sin^{2a+1}(\theta) \; \md \theta 
	= \left\{ 
		\begin{array}{l l}
  		0 & \quad \text{if $k+q$ is odd}\\
  		2\sum_{p=0}^a {a \choose p} \frac{(-1)^p}{2(p+K)+1} \equiv C^a_K & \quad \text{otherwise, $k+q=2K$}
  		\end{array}
  	   \right.
\end{align}
Thus we have for now 
\begin{align}
\label{general_spin_full_equation}
&\trace{\rho_{\cS}^{jj'}(t) \rho_{\cS}^{j'j}(t)} 
	= (2\pi)^2 \sum_{a=0}^{2j}{2j \choose a}{2j' \choose a} \nonumber \\
		&\left[ \sum_{k+q= 2K}^{2j-a,2j'-a}
		{2j-a \choose k } {2j'-a \choose q} (-1)^K C_K^a
		\int_0^{\infty} \cos^{2(j+j')-2(K+a)}\left(\frac{pt}{2}\right) 
		\sin^{2(K+a)}\left( \frac{pt}{2} \right)
		 p^2 \me^{-2\lambda  p^2}
		\right]^2
\end{align}
Only remains now the integral over the norm variable $p$ where all the time dependance is still hidden. 
In fact we are only interested on the asymptotic behavior in time of the norm in order to conclude on 
decoherence of the superposition or not. The important result is that we have a zero limit only on the 
case when $2(j+j')$ is an odd number, meaning that we have initially an integer/half-integer superposition.
Otherwise we have a non zero limit. From the integral
\begin{align}
\lim_{t \rightarrow \infty} 
	\int_0^{\infty} \cos^{2(j+j')-2(K+a)}\left(\frac{pt}{2}\right) 
		\sin^{2(K+a)}\left( \frac{pt}{2} \right)
		 p^2 \me^{-2\lambda  p^2} 
	= \frac{\sqrt{\pi}}{4^{j+j'+1}} \frac{{2(j+j') \choose j+j'}{j+j' \choose a+K}}{{2(j+j') \choose 2(a+K)}}
\end{align}
Putting everything together, we finally have 
\begin{align}
\label{limit_norm}
\lim_{t \rightarrow \infty} \trace{\rho_{\cS}^{jj'}(t) \rho_{\cS}^{j'j}(t)} 
	= \left(\frac{2\pi}{\cN'}\right)^2 \frac{\pi}{4^{2(j+j'+1)}} {2(j+j') \choose j+j'}^2
		\sum_{a=0}^{2j}{2j \choose a}{2j' \choose a}
		\left[ 
		\sum_{k+q= 2K}^{2j-a,2j'-a}
		(-1)^K 
		\frac{{2j-a \choose k } {2j'-a \choose q}{j+j' \choose a+K}}{{2(j+j') \choose 2(a+K)}}
		C_K^a
		\right]^2
\end{align}
This limit does not always vanish and we give a table of those limits for low spins:
\begin{table}[h!t]
\begin{center}
\begin{tabular}{|l|c|c|c|c|c|c|}
  \hline
\diagbox[width=3em]  {j}{j'}
  & 0 & 1/2 & 1 & 3/2&2&5/2 \\
  \hline
  0 & $\pi^3$ & 0 & $\pi^3/9$ & 0 & $\pi^3/25$  & 0 \\
  \hline
  1/2& 0  & $2\pi^2/9$ & 0 & 1.34 &  0 &  0.65 \\
  \hline
  1 & $\pi^3/9$ &0 & 2.61 & 0 & 0.67 & 0    \\
  \hline
  3/2 & 0 & 1.34 & 0 & 1.31 &0 & 0.4  \\
  \hline
  2 & $\pi^3/25$ & 0& 0.67 &0 & 0.78 & 0\\
  \hline
\end{tabular}

\end{center}
\caption{Some numerical values of the limit of the coherence at infinity. Each value is a rational number
that can be obtained by evaluating formula \eqref{limit_norm}. A coherence remains only for integer
or half-integer superposition.}
\label{tableau_particules}
\end{table}

\begin{figure}[!ht]
\begin{center}
\includegraphics[width=\textwidth]{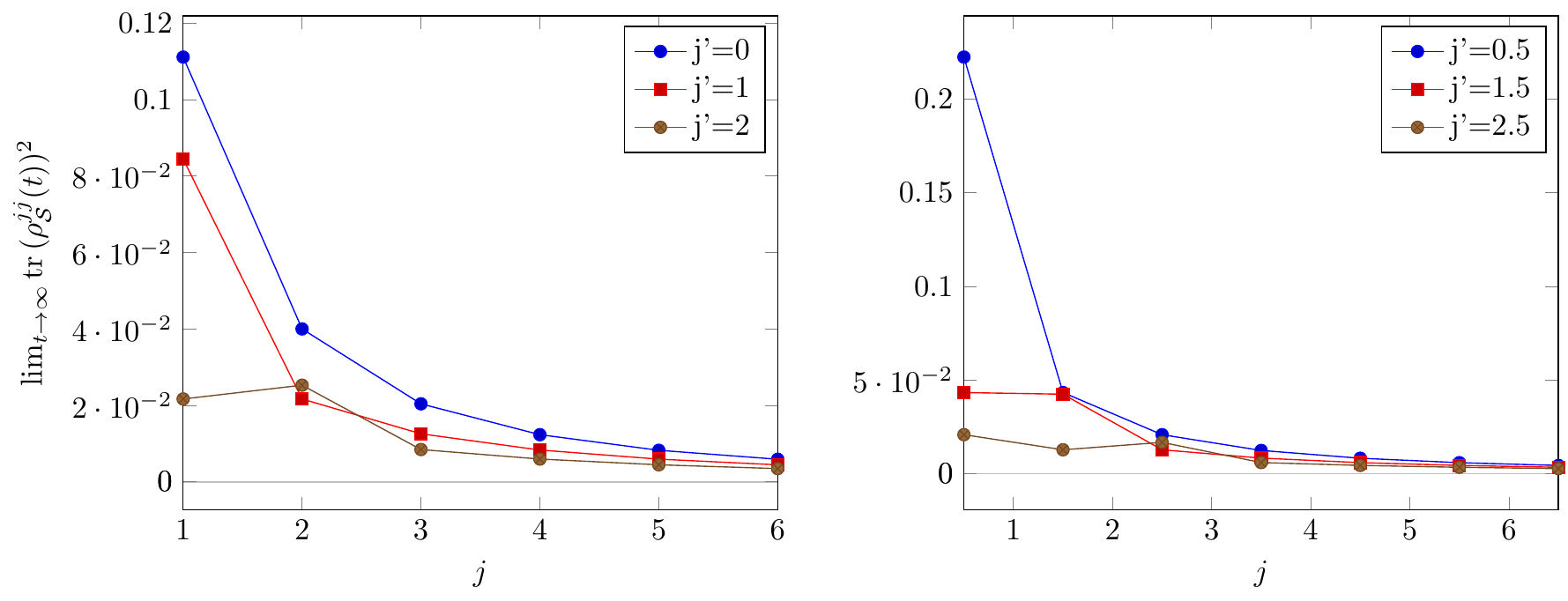}
\caption{Some numerical values of the limit of the coherence at infinity. Each value is a rational number
that can be obtained by evaluating formula \eqref{limit_norm}. A coherence remains only for integer
or half-integer superposition but tends rapidly to zero as the spins get higher.
 \label{limits_numerical}}
\end{center}
\end{figure}

This proves the statement that there exists some type of  coherence of superposition  as time 
goes to infinity. It is a straightforward check to recover from this generic formula the limits in the $j'=0$
case given by eq.\eqref{limit_j'_0}. Fig \ref{limits_numerical} shows the general behavior
of the remaining coherence as a functions of the spins. This remaining coherence at late time $t\rightarrow \infty$ decreases as the spins $j$ and $j'$ grows. So in a semi-classical regime for large quanta of area, we could conclude for an almost-total decoherence. But in the deep quantum regime, with Planck size excitations of the geometry, this remaining coherence might play a non-trivial role.

\subsection{On the decoherence timescale}

The form of the projected reduced density matrix can be obtained exactly at any time from eq.\eqref{general_spin_full_equation} 
and fig.\ref{limits_numerical} represents its typical behavior in the two distinct cases of a boson-boson type superposition and a 
boson-fermion type superposition. Since the former has a non zero limit as time goes to infinity,
a coherence always remains between those states.

The short timescale behavior is dominated by a Gaussian decay with a damping time inversely 
proportional to the ``squared distance'' between the spins $(j-j')^2$ but also to their sum
$j+j'$. This evolution can be obtained by a straightforward expansion at leading order in the  time $t$ of eq.\eqref{fullequation},
\begin{align}
\label{exact_expansion}
\trace{\rho_{\cS}^{jj'}(t) \rho_{\cS}^{j'j}(t)} \underset{t\rightarrow 0}{\simeq}
	\trace{\rho_{\cS}^{00} \rho_{\cS}^{00}}
		\left( 1 - [(j+j') + (j-j')^2] \right) \frac{t^2}{4}
\end{align}

This suggest at first sight a decoherence between states with different spins. What's more,
the state most immune to the interaction with the environment is the rotation invariant $j=0$ state
as suggested by the $j+j'$ damping. This is natural in the light of the interaction which couples the three 
rotation operators $L_i$ to the environment.

However, on the long run, a re-coherence appears in the superposition and different conclusions 
must be drawn. Re-coherence is a natural phenomenon when a finite size environment (with all free
dynamics taken into account) is considered. The associated timescale depends on the number of modes 
of the  environment. Only in the limit of an infinite size environment can we obtain a true decoherence 
at all time but for all practical purposes the timescale can be extremely long.

As stated previously, for the problem at hand, a coherence remains between superposition 
of two integer or two half-integer spins. The limit value of the norm depends of course on the spins of 
the superposition. For instance we had the scaling law in $1/(2j+1)^2$ for the case with $j'=0$.
For integer/half-integer spins superposition the coherence dies out as time goes to infinity but with a
typical timescale completely independent of the spins. This can be straightforwardly obtained by 
expanding equation \eqref{general_spin_full_equation} : the asymptotic behaviors are Gaussian
of the form $\me^{-\frac{t^2}{8}}$ or $\me^{-\frac{1}{4}\frac{t^2}{8}}$ respectively 
from the integer superposition and the half-integer superposition.

Finally, if we consider a non-dynamical bath of oscillators for the environment, all those coherence 
times and limits are rescaled by the number $N$ of oscillators given by formula $\eqref{densitymatrix_bath}$.

\begin{figure}[!ht]
  \label{timescales}
\begin{center}
\includegraphics[width=\textwidth]{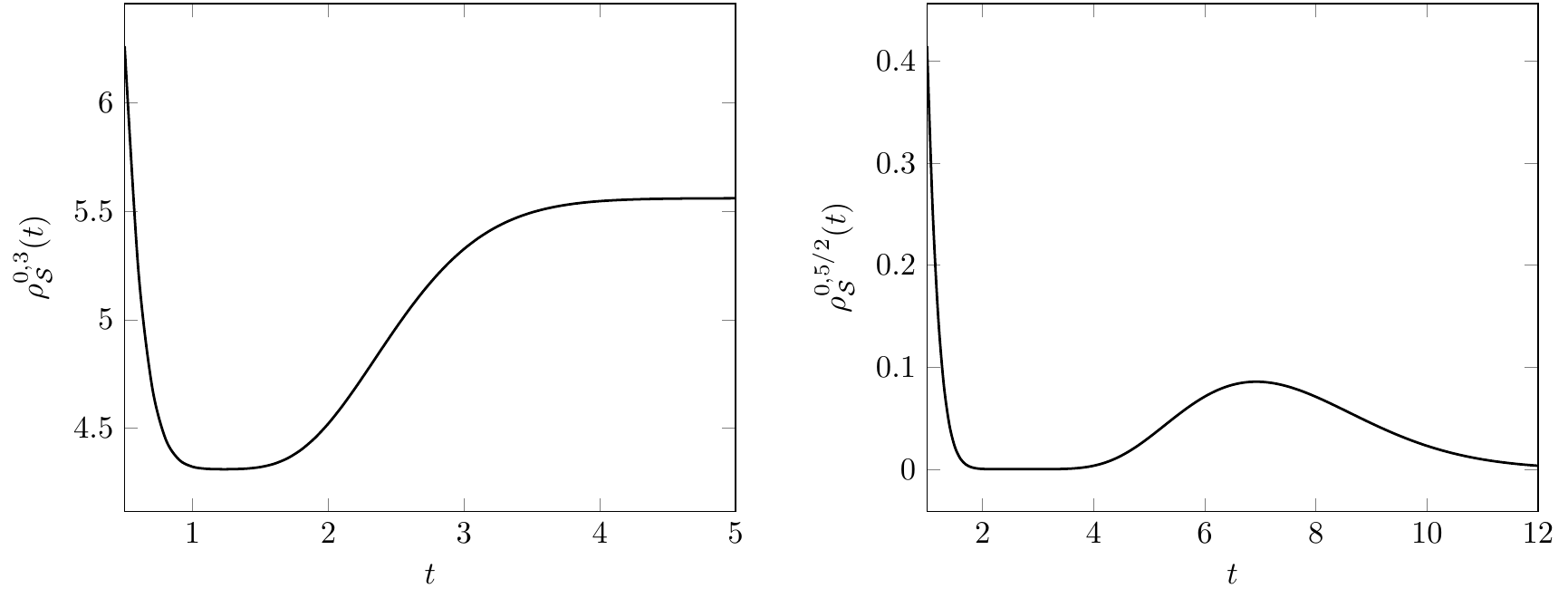}
\caption{Numerical evaluations of the two typical behavior of the norm of the spin coherence (the beginning is Gaussian 
		and has been omitted in the plots to highlights the non trivial structures). At first
 		the coherence tends to diminish. However a re-coherence occurs and depending on 
 		the nature of the superposition the coherence saturates to a non-zero value (boson/boson 
 		like superposition) or tends to zero (fermion/boson like superposition).}
\end{center}
\end{figure}

\subsection{A natural extension : how to get rid of recoherence}

The analysis of the $N=2$ patches model shows that the interaction as it is does not lead to
decoherence on states with a definite closure defect value. The pointer states are not 
eigenstates of the operator $\mathbf{L}^2$. A natural solution to this problem is to force
the environment to couple to this operator. 

Let's consider the formal interaction between $L=|\mathbf{L}|$ and $p = |\mathbf{p}|$, adding a term $L\otimes p$ to the original interaction Hamiltonian $H_{\cS \cE}$ that we postulated in \eqref{N2_hamiltonian}.
The analysis is quite simplified by the fact that this term in the Hamiltonian commutes with the first, 
so we can consider it alone. It leads to a decoherence between different spins. 
Consider again the superposition \eqref{superposition}. Since $L\ket{j,g} = j\ket{j,g}$, 
the superposition evolves at time $t$ into the state
\begin{align}
\ket{\psi_{\cS\cE}(t)} = \frac{1}{\sqrt{\cN}} 
				\left( 
				\ket{j,g} (\me^{-\frac{\mi t}{\hbar} j p} \ket{0}) +
				\ket{j',g'} (\me^{-\frac{\mi t}{\hbar} j' p} \ket{0})
				\right)			
\end{align}
The states of the environment have the from $\ket{E_j (t)} = (\me^{-\frac{\mi t}{\hbar} j p} \ket{0}) $.
The decoherence factor for the non-diagonal matrix elements of the reduced density
for the system is then the overlap $\langle E_{j'}(t) | E_j(t) \rangle$. A straightforward calculation
in the momentum basis gives the explicit Gaussian behavior
\begin{align}
\langle E_{j'}(t) | E_j(t) \rangle = \me^{-(j-j')^2 t^2} = \me^{-\frac{t^2}{\tau^2_{\text{deco}}}}	
\end{align} 
with the decoherence time $\tau_{\text{deco}} = 1/|j-j'|$. Their is thus a decoherence for 
spin superposition with a damping time inversely proportional to the distance between the spins.

The operator $\mathbf{L}^2$ can be written in terms of deformation operators. Using the relations
$\mathbf{L}_i.\mathbf{L}_j = \frac{1}{2}E_{ij}E_{ij}^{\dagger}-\frac{1}{4}E_iE_j-\frac{1}{2}E_i$
or in terms of the creation operators $F_{ij} = a_i b_j - a_j b_i$,  
$\mathbf{L}_i.\mathbf{L}_j = \frac{1}{4}E_i E_j-\frac{1}{2} F_{ij}^{\dagger}F_{ij}$, we can obtain
the relations 
\begin{align}
\mathbf{L}^2 = \frac{E}{2}\left( \frac{E}{2} +1\right) - F_{12}^{\dagger}F_{12}
		= \frac{1}{4}(E_1-E_2)^2 + \frac{1}{2}(E_{12}^{\dagger}E_{12} + E_{21}^{\dagger}E_{21})
\end{align} 
Those operators could be coupled to the environment to induce a decoherence on the closure defect.
In essence it amounts to couple the Casimir operator \eqref{Casimir_U(N)} to the environment.

To conclude the discussion of the specific surface state with $N=2$ patches with a non dynamical 
environment, we have shown that the dynamic induces a decoherence effect for superposition of different
spins on short timescales. A re-coherence occurs on the long run and we concluded on the damping of coherence
only for superpositions of integer/half-integer spins. If we insists on having a decoherence on the closure
defect, a natural solution is to introduce a new coupling to the environment.

\section{Master equation approaches}
\label{Master equation approaches}

Most models of open quantum systems and studies of decoherence  are not exactly solvable and 
approximate methods have to be developed. Master equations based on Born-Markov approximations
are the ones most commonly used for analyzing open quantum dynamics in quantum optics and 
condensed matter physics. They are equations for the reduced density matrix of the system taking 
into account the effects of the environment to first order. They are relevant for understanding the 
behavior of the system at a time $t$ much longer than any correlation times $\tau_c$ but still shorter
than dynamical timescales $T$ : $\tau_c \ll t \ll T$. This is the essence of the Markov approximation.
A large environment is needed to neglect the changes of the state of the environment due to 
the coupling to the system and correlations up to second order. 

In the following, we apply the master equation methods to the problem of open quantum surface dynamic
by first deriving the Born-Markov master equation. This step will motivate a more phenomenological
approach by postulating jump operators for the Lindblad equation. The results of those different approaches are 
then compared to the exact results obtained previously.

\subsection{Born-Markov equation} 

An approximate equation for the reduced density matrix  of the system can be derived by an expansion 
of the exact equation of motion 
$\frac{\md\rho(t)}{\md t}  =  -\frac{\mi}{\hbar} [H, \rho(t) ]$  \cite{Haroche_book}. 
For an interaction written  has $H_{\cS\cE} = \sum_i S_i \otimes E_i$. It has the general form
\begin{align}
\label{redfield}
\frac{\md\rho_{\cS}(t)}{\md t}  = 
	-\frac{\mi}{\hbar} [H_{\cS}, \rho_{\cS}(t) ]
	+\frac{1}{\hbar^2} \sum_i U_i(t) \rho_{\cS}(t) S_i + S_i \rho_{\cS}(t) U_i^{\dagger}(t)
	-S_i U_i(t) \rho_{\cS}(t) - \rho_{\cS}(t) U_i^{\dagger}(t) S_i
\end{align}
with the operators $U_i(t) = \int_0^{t} \sum_j g_{ij}(\tau)S_j(-\tau) \; \md \tau$ encoding the action 
of the environment on the system and depend on its correlation functions 
$g_{ij}(\tau)=\langle E_i(t)E_j(t-\tau)\rangle_{\rho_{\cE}}$. 

To go further, the behavior in time of the correlation functions must be discussed. It depend 
naturally on the proper dynamic of the environment $H_\cE$ and on the state $\rho_\cE$.
For a dynamical environment, the correlation functions decay over a timescale $\tau_c$ 
called correlation time or memory time. 
Denoting by $v$ an order of magnitude of an element of matrix of the interaction, equation
\eqref{redfield} is an expansion in the parameter $v\tau_c/\hbar$.
The order of magnitude of the coupling in the Born-Markov equation is $v^2\tau_c/\hbar$ which is then 
much smaller than the memory frequency $\tau_c^{-1}$ in the short memory 
time approximation. The complete Born-Markov equation is then obtained by 
approximating the integral in $U_i(t)$ by its value at infinite time giving in the end a pure 
local in time equation of motion. However, if the environment were small or non dynamical, the natural
expansion parameter would be $vt/\hbar$ and the results of the Born-Markov equation would 
be inaccurate on long timescales and the time dependence of the correlation functions must be kept. 
This is an issue we will discuss further in the section comparing the different approaches.

Now for the specific problem we are interested in, we use the interaction \eqref{interaction} 
and express the Born-Markov equation. The equation  is here simplified by the fact 
that we neglect the proper dynamic of the surface. In particular, the operators $U_{ij}$
have the simple from $U_{ij} = \sum_{kl} \left(  \int_0^{t} g_{ij,kl}(\tau) \; \md \tau \right)  E_{kl}$. 
After some straightforward algebra using the $U(N)$ commutation relations, we have the equation 
\begin{align}
\label{BM}
\frac{\md}{\md t} \rho_{\mathcal{S}}(t) = - 
	&\sum_{ijk} E_{ij} 
		\rho_{\mathcal{S}}(t) 
		\int_{0}^{t} \left( g_{ik,kj}(\tau) - g_{ki,jk}(\tau) \right) \; \md \tau
		+ \rho_{\mathcal{S}}(t) E_{ij} 
		\int_{0}^{t} \left( g_{ik,kj}(-\tau) - g_{ki,jk}(-\tau) \right) \; \md \tau + \\ 
	&\sum_{ijkl} E_{kl}\left[E_{ij}, \rho_{\mathcal{S}}(t)\right] 
		\int_{0}^{t} g_{ij,kl}(\tau) \; \md \tau
		+ \left[\rho_{\mathcal{S}}(t), E_{ij} \right] E_{kl} 
		\int_{0}^{t} g_{kl,ij}(-\tau) \; \md \tau\notag
\end{align}
To go further we have to specify the form of the correlation functions. First it is natural to expect
the correlation function to be symmetric in time. To be more specific, let's imagine we have an 
harmonic environment and that the operator $V_{ij}$ creates a photon at $j$ with creation operator
$\gamma_j^{\dagger}$  (a quanta of area is destroyed) and absorbs one at $i$ with destruction 
operator $\gamma_i$ (a quanta of area is created) so $V_{ij} = \gamma_i^{\dagger} \gamma_j$.
For the environment in the vacuum state or thermal state
(any Gaussian states), the Wick theorem applies and allows to develop the correlation functions.
Replacing those correlation functions into the master equation is then straightforward. For an 
isotropic, homogeneous non dynamical environment, we obtain the simplest form of the equation
\begin{align}
\label{BM_Lindblad}
\frac{\md\rho_{\mathcal{S}}}{\md t}
= \sum_{i,j =1}^{N} \kappa(t)
	\left[ 
	E_{ij}\rho_{\mathcal{S}}E_{ij}^{\dagger}
	- \frac{1}{2}
	\left( 
	E_{ij}^{\dagger}E_{ij} \rho_{\mathcal{S}} + \rho_{\mathcal{S}}E_{ij}^{\dagger}E_{ij}  
	\right)
	\right]
\end{align}
where $\kappa(t) = t \kappa$ with $\kappa$ a constant function of the correlation function. This master equation 
has the Lindblad form. In the full Born-Markov approximation, $\kappa(t)$ would be independent of time and 
a decoherence would be expected a priori with an exponential decay $\me^{-t/\tau_d}$ with $\tau_d$ a 
decoherence timescale. Here however, the linear time dependence caused by the non dynamical 
character of the environment (non markovianity) leads to a decoherence with a Gaussian behavior $\me^{-t^2/\tau_d^2}$. 
This form is in full agreement with the short time exact calculations \eqref{exact_expansion}.

\subsection{Lindblad approach} 

Once again, we focus on the simplest $N=2$ patches model
with the spin interaction part and take a phenomenological approach to it with the Lindblad master
equation. The jump operators (Lindblad operators) are the spin $L_i$ operators and no free dynamic  
is supposed to occur for the system. We should not forget that we really consider the Schwinger 
representation here and that we work not in the Hilbert space at a given spin $j$. Superposition of 
states with different values of the spin $j$ are permitted. 
Let's emphasize some subtleties regarding the correlation functions and the definition of the jump 
operators in order to compare those master equation approaches to the exact dynamic proposed 
in the last section due to the hypothesis of a non-dynamical environment $H_\cE = 0$. We keep in mind 
this important point but discuss now in a phenomenological way a Lindblad equation with $J_i$ 
jump operators as done traditionally in quantum optics models.

The master equation we propose to study is thus
\begin{align}
\frac{\md\rho_{\mathcal{S}}}{\md t} 
= \sum_{i = x,y,z}
	L_i\rho_{\mathcal{S}}L_i -\frac{1}{2}\left( L_iL_i\rho_{\mathcal{S}} + \rho_{\mathcal{S}}L_iL_i \right)
= \sum_{i = x,y,z} L_i\rho_{\mathcal{S}}L_i -\frac{1}{2}\left( \mathbf{L}^{2}\rho_{\mathcal{S}}  
	+ \rho_{\mathcal{S}} \mathbf{L}^{2} \right)
\end{align}
For the surface dynamic we are ultimately interested in, we want to understand if their is a decoherence
phenomena on superposition with different values of the spin $j$. Since $\mathbf{L}^{2}$ commutes 
with the jump operators, the environment does not induce transitions between states with different
spins and no dissipation occurs. To focus on coherence between different spin states, we can look again at the 
projection of the reduced density matrix $\rho_{kl} = P_k \rho P_l$ with $P_{k,l}$ the projection 
operator on the subspace of spin $k$ and $l$ respectively.
\begin{align}
\frac{\md\rho_{kl}}{\md t} = 
\sum_{i = x,y,z} L_i\rho_{kl}L_i -\frac{1}{2}
\left( k(k+1) + l(l+1) \right)\rho_{kl} \notag 
\end{align}
Searching for pointer states (approximate pointer states generally) requires to evaluate
an entanglement witness such as the Von Neumann entropy or the purity of the states\footnote{
The choice of one or the other should in a proper limit gives the same approximate results.}.
For our purpose we will mostly focus on the norm of the projected reduced density matrix
and its evolution.
\begin{align}
\frac{\md \trace{\rho_{kl}\rho_{kl}^{\dagger}}}{\md t}  = \sum_{i = x,y,z} 
2\trace{L_i\rho_{kl}L_i\rho_{lk}} - \left( k(k+1) + l(l+1) \right)\trace{\rho_{kl}\rho_{lk}}
\end{align}
Let's for instance look at the short time evolution of the superposition 
 $\ket{\psi} = \frac{\ket{k,k} + \ket{l,l}}{\sqrt{2}}$,
\begin{align}
 \left.\frac{\md \trace{\rho_{kl}\rho_{kl}^{\dagger}}}{\md t}\right|_{t=0}
 = \left[ 2kl-\left(k(k+1) + l(l+1)\right) \right] \trace{\rho_{kl}\rho_{kl}^{\dagger}}
 = -\left[ \left(k-l\right)^2 + (k+l)\right] \trace{\rho_{kl}\rho_{kl}^{\dagger}}
\end{align}
We thus qualitatively see that a superposition of different spin states leads to a 
more rapid entanglement with the environment than a state with definite spin. 
Moreover we see that only a rotation invariant state is immune to entanglement (at 
first order) with the environment whereas even a state with a definite spin gets entangled
with its environment (the higher the spin the more entangled). This behavior can be 
generalized to an arbitrary initial pure state (for the proof see appendix \ref{proof})
\begin{align}
\left.\frac{\md \trace{\rho_{kl}\rho_{kl}^{\dagger}}}{\md t}\right|_{t=0} \leq
-\left[\left(k-l\right)^2 + (k+l) \right]\trace{\rho_{kl}\rho_{kl}^{\dagger}} 
\end{align}

 Let's discuss now the relations between the different approaches and in 
particular why the conclusions appear not to be the same. We have explored in two different ways 
a possible decoherence effect for quantum superposition of states with different 
spins associated to the closure defect. The first method was based on an exact calculation
for the $N=2$ patches model and the second used the traditional methods of Markovian
master equations.

\begin{itemize}

\item  The ingredient for master equations to work is to have a large enough dynamical environment
for its correlation functions to vanish on a timescale smaller than any relaxation or observational times.
Qualitatively said, the environment is without memory. In this context, we have shown that the 
surface (approximate) pointer states are those with a given value of the closure defect and that 
the decoherence factor has an exponential decay $\me^{-t/\tau_d}$.  

The behavior  predicted by the exact approach on a short-time scale eq.\eqref{exact_expansion} is in fact Gaussian.
This is easily understood when remembering that the exact dynamic was studied for a non-dynamical
environment which then acts as a classical fluctuating potential.  The integrals in \eqref{BM} cannot be extended 
to infinite time. We thus have memory effects and a linear dependance in time in $\kappa(t)$ and 
so linear time dependent jump operators. The Born-Markov analysis eq.\eqref{BM}\eqref{BM_Lindblad} would 
only be meaningful on short time-scale and would naturally lead to Gaussian decay functions 
$\me^{-t^2/\tau_d^2}$. The difference between Gaussian and exponential decay is thus traced
back the memory of the environment controlled by its dynamic.

\item The predictions on the decoherence effect differ for the two methods and only match on a 
short timescale. In particular the exact analysis shows that a recoherence occurs with a non zero 
limit  (a limit still approaching zero as the spins get higher). If as expected the closure defect is associated to 
a quasi-local energy density and the curvature or torsion it generates, the spin is also expected to 
be high enough for a black hole. Thus for all practical purposes, we can conclude on an effective decoherence
on the closure defect.

\end{itemize}

\section{Conclusion}
\label{Conclusion}

Decoherence is a now a cornerstone of quantum physics to clarify the quantum 
to classical transition. In a theory of quantum gravity, the geometry is a dynamical
and fluctuating field and quantum superposition of geometry are perfectly allowed states.
Their non observability in the classical regime remains to be clarified in the semi-classical 
analysis of loop quantum gravity. Our first investigation focus on the open dynamic 
of a quantum surface coupled to an environment comprising all the other gravitational
and matter degrees of freedom of the Universe. This bulk-boundary coupling induces 
a decoherence and our aim was to understand the emergence of some geometrical
super-selection sector. 

Through the deformation formalism of quantum geometry, we proposed toy models 
for the open dynamic of a quantum surface in the context of loop quantum gravity and
a natural coupling between a bath of harmonic oscillators (modeling for instance quantum matter 
fields) and the deformations of the surface.
We looked for a decoherence on the closure defect of a surface with fixed area using
two different methods: one exact method analyzing the physical effect on a superposition
of the interaction part of the Hamiltonian (quantum measurement limit) and the other 
using master equations approaches under Born-Markov approximations. The two 
approaches agree on the short timescale and indeed conclude on a decoherence of quantum 
superposition of states with different spins associated to the closure defect. The decoherence
factor is here a Gaussian decaying with a timescale inversely proportional to the spin 
difference. However due to the different treatment of the structure of the environment, 
the conclusions differ as time goes to infinity. The exact treatment neglects the proper dynamic 
of the environment which thus has an infinite correlation time (constant correlation functions) 
and leads to a re-coherence of integer/integer or half-integer/half-integer superpositions.
Nonetheless this non zero limit is for all practical purposes irrelevant when large spins 
are considered which is potentially the case for black holes since closure defect
should be a sign of the presence of quasi-local energy in the region that induces
curvature and torsion.

\begin{figure}[!ht]
  \label{BlackHole}
\begin{center}
	\begin{picture}(200,180)
		\put(0,0){\includegraphics[width=0.4\textwidth]{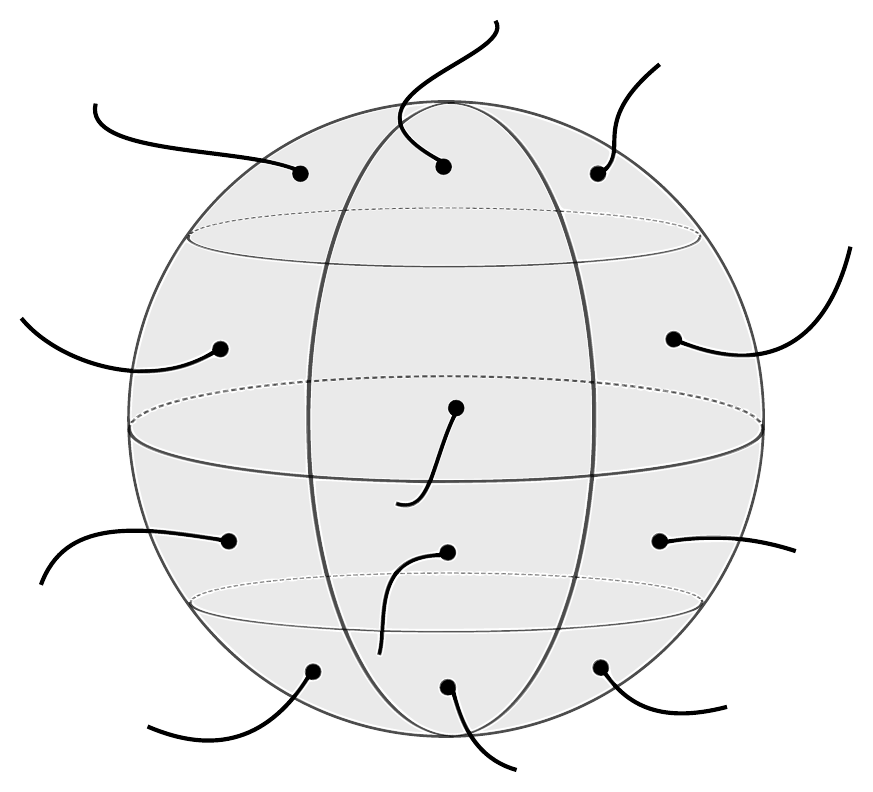}}		
		\put(90,110){$\phi_{\text{bulk}}$}
	\end{picture}
 \caption{Section of an horizon in loop quantum gravity seen as a punctured surface. The bulk of the inside
 of the black hole is filled with matter fields and gravity degrees of freedom denoted generically $\phi_{\text{bulk}}$
 hidden \emph{a priori} to the outside  observer. The induced dynamic of the surface can be non 
 trivial like a Bose-Hubbard like model.}
\end{center}
\end{figure}

From the present construction, this surface dynamic model and the study of 
decoherence can be refined along different lines

\begin{itemize}

\item The free dynamics can be properly taken into account. This would allow for instance to rigorously 
verify that an environment without memory would lead to a full decoherence on the spins since 
the compactness of the $\SU(2)$ group could not be seen by the environment. 

\item A drawback of the current approach is that we are considering a dynamic and a decoherence of a 
geometry evolving in a given classical time. To be more true to the relativistic point of view, it would be 
most interesting to have a quantum model where a classical notion of time would emerge from 
a decoherence process along with the decoherence on geometric properties. We would 
then look at the flow of correlations between two observables of a system and a quantum clock. 
Some relationships between an intrinsic decoherence induced be a (discrete) quantum time 
have been explored in \cite{Milburn_1991,Milburn_2003}.

\item Before considering even the coupling of the boundary surface and the bulk, and 
instead of the natural harmonic oscillators dynamic for the system, we could consider a more involved 
model for the free boundary such as a Bose-Hubbard model. The horizon of the black hole 
would then be seen as an interacting gaz of punctures \cite{Perez_2014, Noui_2015}.
The phase diagram as a function of the mass and 
temperature could then be studied, checking that at high mass their exists a superfluid phase and Bose
gaz phase at small mass respectively characterized by a diffusive and a ballistic response to local
perturbations.

\end{itemize}

The semi-classical analysis of loop quantum gravity has mostly up to now been focused
on understanding coherent states interpolating a quantum and classical geometry and on 
the coarse-graining of spin network states. Still, it is an important and non trivial issue 
to understand in a quantum theory of gravity the quantum to classical transition through 
a decoherence mechanism and poses some conceptual questions.
From the perspective of describing quantum gravity from quantum information, for instance
computing entanglement entropy and decoherence effects, a proper definition on the 
separation between bulk, boundary and exterior degrees of freedom in quantum gravity 
has to be found \cite{Freidel_2016}. The subtleties come from the gauge invariance or 
diffeomorphism invariance. The state of an exterior observer is then obtained by tracing out 
the bulk degrees of freedom composed of matter and gravitational degrees of freedom.
This step raises questions again in light of the holographic principle from which we learn 
that the bulk degrees of freedom are fully encoded on the boundary. The very meaning
of tracing out bulk degrees of freedom is quite ambiguous. After clarifying those conceptual
issues, we could then investigate the existence of some decoherence phenomena seen 
by an outside observer on the horizon induced by the bulk-boundary coupling and identify 
the semi-classical states (pointer states) selected by the bulk or better understand 
the relationship between coarse-graining methods and tracing out degrees of freedom.

\section*{Acknowledgments}

We would like to thank Nad\`ege Lemarchand for her numerical study of the coherence evolution in our toy model in the context of her Masters internship (2015) at the ENS de Lyon.

\appendix

\section{Reduced density matrix of the environment}
\label{Density_matrix_env}

In the core of this paper, we analyzed the reduced density matrix of the surface in 
contact an unmonitored environment. We could also look at the behavior of the reduced 
density matrix of the environment. We recall that the environment is composed of all the degrees
of freedom (matter...) in the Universe except those associated to the system.

Consider the state
\begin{align}
\ket{\psi_{\cS\cE}} = \frac{1}{\sqrt{\cN}} \ket{j,g} \otimes 
				\left(\ket{\mathbf{p}}+\ket{\mathbf{q}} \right)
\end{align}
Since we have 
$U(t)\ket{j,g}\ket{\mathbf{p}} = \ket{j, \me^{-\frac{\mi t }{\hbar}\frac{\boldsymbol{\sigma}.\mathbf{p}}{2}} g}
\ket{\mathbf{p}}$, the state at time $t$ is simply
\begin{align}
\ket{\psi_{\cS\cE}(t)} = \frac{1}{\sqrt{\cN}} 
				\left(\ket{j,D(tp/2,\hat{\mathbf{p}}) g}\ket{\mathbf{p}} +
					\ket{j,D(tq/2,\hat{\mathbf{q}}) g}\ket{\mathbf{q}} \right)
\end{align}
Tracing over the surface, we can obtain the reduced density matrix of the environment. The 
coherence terms are modulated by the decoherence factor which is the overlap 
\begin{align}
\langle j, D(tq/2,\hat{\mathbf{q}}) g | j,D(tp/2,\hat{\mathbf{p}}) g \rangle =
\langle D(tq/2,\hat{\mathbf{q}}) g | D(tp/2,\hat{\mathbf{p}}) g \rangle ^{2j}
\end{align}
We specify the calculation to the spin up case $g=\begin{pmatrix}1 \\ 0 \end{pmatrix}$ to have
an explicit form of the overlap 
\begin{align}
\label{deco_factor_env}
\langle D(tq/2,\hat{\mathbf{q}}) g | D(tp/2,\hat{\mathbf{p}}) g \rangle ^{2j} = 
	&\left(
		\cos(tp/2)\cos(tq/2) + \sin(tp/2)\sin(tq/2) \hat{\mathbf{p}}.\hat{\mathbf{q}} \right. \nonumber \\
		&-\left.\mi(\cos(tp/2)\sin(tq/2) \hat{\mathbf{q}}.\boldsymbol{\sigma}
			+ \cos(tq/2)\sin(tp/2) \hat{\mathbf{p}}.\boldsymbol{\sigma}
			+\sin(tp/2)\sin(tq/2) (\hat{\mathbf{p}}\wedge \hat{\mathbf{q}}). \boldsymbol{\sigma})
	\right)^{2j}
	\nonumber
\end{align}
A more general state for the system could be considered as a superposition on the spins 
$\sum_j \alpha_j \ket{j,g}$, thus generalizing the overlap \eqref{deco_factor_env} to 
\begin{align}
O_{\mathbf{p}\mathbf{q}}(t) =
\sum_j |\alpha_j|^2
\langle D(tq/2,\hat{\mathbf{q}}) g | D(tp/2,\hat{\mathbf{p}}) g \rangle ^{2j}
\end{align}
Let's consider a particular superposition with amplitude $\alpha_j = 1/\sqrt{2j!}$. This simplify 
the overlap to an exponential 
\begin{align}
O_{\mathbf{p}\mathbf{q}}(t) =
\me^{
\langle D(tq/2,\hat{\mathbf{q}}) g | D(tp/2,\hat{\mathbf{p}}) g \rangle }
\end{align}
The phase of this overlap is correspond to some relaxation whereas the modulus is the
decoherence factor $\cD_{\mathbf{p}\mathbf{q}}(t)$ of the superposition that has the 
simple form 
\begin{align}
\label{deco_time_env}
\cD_{\mathbf{p}\mathbf{q}}(t) = \me^{\cos(tp/2)\cos(tq/2) + \sin(tp/2)\sin(tq/2) \hat{\mathbf{p}}.\hat{\mathbf{q}}}
\end{align}
This decoherence factor is periodic in time and thus does not lead to a proper decoherence in the 
momentum as one could have expected from the interaction form. This origin of this periodicity 
can be traced back the compact structure of $\SU(2) \sim S^3$. 

Let's show that for small time, we recover a decoherence in the momentum comparable to the one 
obtained in ``the flat case interaction'' $H_{\cS\cE} = \mathbf{x} \otimes \mathbf{p}$. For this interaction, it is
straightforward to show that the decoherence factor has the form
$\cD_{\mathbf{p}\mathbf{q}}^{\text{flact}}(t) = \me^{-(\mathbf{p}-\mathbf{q})^2t^2}$. Doing the expansion 
in time of eq.\eqref{deco_time_env}, we have 
\begin{align}
\cD_{\mathbf{p}\mathbf{q}}(t) \propto \me^{-\frac{(p^2 + q^2)}{2} t^2 + t^2 \mathbf{p}.\mathbf{q}}
					= \me^{-\frac{(\mathbf{p}-\mathbf{q})^2}{2}t^2}
\end{align}
As long as the structure of the rotation group is not explored completely, we obtain the same decoherence
effect as in the flat case. This is a consequence of the local flatness of $\SU(2)$. 

\section{Proof the the bound on the purity evolution}
\label{proof}

\quote{
\textbf{Proposition}: \textit{For an initially pure state of the system, the short time behavior 
of the purity evolve according to} 
\begin{align}
\left.\frac{\md \trace{\rho_{kl}\rho_{kl}^{\dagger}}}{\md t}\right|_{t=0} \leq
-\left[\left(k-l\right)^2 + (k+l) \right]\trace{\rho_{kl}\rho_{kl}^{\dagger}} 
\end{align}
}

\begin{proof}
We want to obtain a differential inequality on the norm of the coherence for the reduced
density matrix of the system. 
Consider then a pure state $\ket{\psi}$ of the system and develop it on the coherent states basis. 
\begin{align*}
\rho_{\mathcal{S}} = \ket{\psi} \bra{\psi} \quad 
\ket{\psi} = \sum_j \int_{S^2} \psi_j(\hat{n}) \ket{j,\hat{n}} \;\md^2n
\end{align*}
With this decomposition, the projected reduced density matrix and its norm are 
\begin{align}
\rho_{kl} &=  \int_{S^2} 
\psi_k(\hat{n}) \psi^*_l (\hat{m}) \ket{k,\hat{n}} \bra{l,\hat{m}} \;\md^2n\md^2m \notag 
\\
\trace{L_i\rho_{kl}L_i\rho_{lk}} &= \int_{S^2} 
\psi_k(\hat{n}) \psi^*_l (\hat{m})  \psi_l(\hat{m}') \psi^*_k (\hat{n}') 
\langle k,\hat{n}'| L_i | k,\hat{n}\rangle \langle l,\hat{m}| L_i |l,\hat{m}'\rangle
\;\md^2n\md^2m\md^2n'\md^2m '
\end{align}
The overlap between two coherent states of the spin operator $L_i$ for arbitrary spin
can be obtained using the special case of the spin $1/2$
\begin{align}
\langle k,\hat{n}'| J_i | k,\hat{n}\rangle = k \langle\hat{n}'| \sigma_i | \hat{n}\rangle
\langle \hat{n}' | \hat{n}\rangle ^{2k-1}
\end{align}
We then write the evolution equation and isolate the contribution we are interested in. 
The aim is then to obtain an inequality on the remaining term. 
\begin{align}
\label{Equa_evolu_coherence}
\frac{\md \trace{\rho_{kl}\rho_{kl}^{\dagger}}}{\md t} = 
&-\left[\left(k-l\right)^2 + (k+l) \right]\trace{\rho_{kl}\rho_{kl}^{\dagger}} 
+2|\alpha_{k}|^2|\alpha_{l}|^2 kl
\int_{S^2} \psi_k(\hat{n}) \psi^*_l (\hat{m})  \psi_l(\hat{m}') \psi^*_k (\hat{n}') 
\langle \hat{n}' | \hat{n}\rangle ^{2k-1}\langle \hat{m} | \hat{m}'\rangle ^{2l-1} \notag \\
&\left( \sum_{i = x,y,z} 
\langle \hat{n}'| \sigma_i | \hat{n}\rangle \langle \hat{m}| \sigma_i | \hat{m}'\rangle 
-
\langle \hat{n}' | \hat{n}\rangle \langle \hat{m} | \hat{m}'\rangle \right)
\;\md^2n\md^2m\md^2n'\md^2m '
\end{align}
Clearly we need to show the integral to be negative to obtain the required result. 
This integral is first of all real. Then by using $\sum_{p = x,y,z} \trace{\sigma_p A \sigma_p B} 
= 2\trace{A}\trace{B} - \trace{AB}$ with $A$ and $B$ two $2\times2$ matrices, we can 
evaluate the sum on the coordinates,  
\begin{align}
 \sum_{i= x,y,z} \langle \hat{n}'| \sigma_i | \hat{n}\rangle 
\langle \hat{m}| \sigma_i | \hat{m}'\rangle =
2\langle n'|m\rangle \langle m | n \rangle -
 \langle \hat{n}' | \hat{n}\rangle \langle \hat{m} | \hat{m}'\rangle
\end{align}
The integral has for now the following form 
\begin{align}
\label{Int_CS}
2\int_{S^2} \psi_k(\hat{n}) \psi^*_l (\hat{m})  \psi_l(\hat{m}') \psi^*_k (\hat{n}') 
\langle \hat{n}' | \hat{n}\rangle ^{2k-1}\langle \hat{m} | \hat{m}'\rangle ^{2l-1}
\left( 
\langle n'|m\rangle \langle m | n \rangle -
\langle \hat{n}' | \hat{n}\rangle \langle \hat{m} | \hat{m}'\rangle
\right)
\;\md^2n\md^2m\md^2n'\md^2m '
\end{align}
The Cauchy-Schwarz inequality will allow us to conclude. To see this, we write the first term
of the integral as a trace
\begin{align}
\int_{S^2} \psi_k(\hat{n}) \psi^*_l (\hat{m})  \psi_l(\hat{m}') \psi^*_k (\hat{n}') 
\langle \hat{n}' | \hat{n}\rangle ^{2k-1} \langle \hat{m} | \hat{m}'\rangle ^{2l-1}
\left(  
\langle n'|m\rangle \langle m | n \rangle
\right)
\;\md^2n\md^2m\md^2n'\md^2m '
 \notag \\
= \trace{
\underbrace{\int_{S^2} \psi_k(\hat{n}) \psi^*_k (\hat{n}')
\langle \hat{n}' | \hat{n}\rangle ^{2k-1} \ket{n}\bra{n'}
\md^2n'\md^2n}_{O_k}
\underbrace{\int_{S^2} \psi_l(\hat{m}') \psi^*_l (\hat{m})
\langle \hat{m} | \hat{m}'\rangle ^{2l-1} \ket{m'}\bra{m}
\md^2m\md^2m'}_{O_l}
}
\end{align}
The two operators $O_k$ and $O_l$ are Hermitians and positives. With
 $0\leq \trace{O_k O_l} \leq \trace{O_l} \trace{O_k}$ we have 
\begin{align}
\trace{O_k O_l} \leq  
\int_{S^2} \psi_k(\hat{n}) \psi^*_k (\hat{n}') \langle \hat{n}' | \hat{n}\rangle ^{2k}
\int_{S^2} \psi_l(\hat{m}') \psi^*_l (\hat{m})\langle \hat{m} | \hat{m}'\rangle ^{2l} 
\end{align}
This conclude the proof that the integral in \eqref{Equa_evolu_coherence} is always 
negative and also the differential inequality we conjectured.
\end{proof}

\bibliographystyle{bib-style}
\bibliography{gravitation,mat_cond,decoherence}

\end{document}